\documentclass[final,5p,times,twocolumn]{elsarticle}
\usepackage{amssymb}
\usepackage{amsmath}
\usepackage{siunitx}
\usepackage{subcaption}
\usepackage{comment}
\usepackage{multirow}
\usepackage{natbib}
\usepackage{lineno}

\newcommand{\um}{\ensuremath{\text{\textmu m}}}

\newcommand{\neqcm}{\mathrm{n_{eq}/cm^2}}

\journal{Nuclear Instruments and Methods in Physics Research Section A}

\begin{document}
\begin{frontmatter}

\title{\boldmath Systematic Investigation of Acceptor Removal in HPK LGADs with Modified Gain Layers}
\author[a]{Yua Murayama}
\author[a]{Mahiro Kobayashi}
\author[c,c-2]{Tomoka Imamura}
\author[b]{Koji Nakamura}
\author[a]{Issei Horikoshi}
\author[a]{Koji Sato}
\author[d]{Masato Terada}
\author[d]{Minoru Hirose}
\author[d]{Tatsuya Masubuchi}
\author[e]{Sayuka Kita}
\author[f]{Hironori Sonobe}
\affiliation[a]{University of Tsukuba,
Tennodai 1-1-1, Tsukuba, Ibaraki, Japan}
\affiliation[b]{High Energy Accelerator Research Organization,
Oho 1-1, Tsukuba, Ibaraki, Japan}
\affiliation[c]{Physikalisch-Technische Bundesanstalt, 
Bundesallee 100, D-38116 Braunschweig, Germany}
\affiliation[c-2]{
Leibniz University Hannover,
Welfengarten 1, 30167 Hannover, Germany
}
\affiliation[d]{
The University of Osaka,
1-3 Machikaneyama-cho, Toyonaka, Osaka, 560-8531, Japan}
\affiliation[e]{
International Center for Elementary Particle Physics, 
The University of Tokyo, Bunkyo-ku, Tokyo 113-0033, Japan}
\affiliation[f]{
Hamamatsu Photonics K.K.,
1126-1, Ichino-cho, Chuo-ku, Hamamatsu City, Shizuoka Pref., 435-8558, Japan
}

\begin{abstract}
Low-Gain Avalanche Diodes (LGADs) are fast silicon sensors with internal charge multiplication and are key candidates for precision timing layers in future high-energy hadron colliders. Their operation in harsh radiation environments, however, is limited by acceptor removal in the gain layer, which reduces the active acceptor concentration and degrades the internal electric field required for avalanche multiplication. Improving the radiation tolerance of the gain layer is therefore essential for future 4D tracking applications.
In this work, we investigated several LGAD prototypes produced in collaboration with Hamamatsu Photonics K.K. (HPK), featuring modified gain-layer designs, including oxygen-modified, carbon-implanted, and boron--phosphorus compensated structures. The sensors were studied after proton and reactor-neutron irradiation. Radiation tolerance was characterized using the acceptor-removal coefficient extracted from I-V measurements and the operation voltage required to recover the timing performance after irradiation.
The results show that carbon implantation is the only approach among those studied here that provides a clear improvement in radiation tolerance. In contrast, neither oxygen-related modification, including the Partially Activated Boron (PAB) approach, nor gain-layer compensation alone yields a significant improvement, and the compensated carbon-implanted structure shows no clear advantage over the carbon-only case. In addition, the acceptor-removal coefficient is found to depend on the irradiation particle type and energy.
\end{abstract}


\end{frontmatter}

\section{Introduction}
Future high-energy and high-luminosity hadron colliders, such as the High-Luminosity Large Hadron Collider (HL-LHC)~\cite{Rossi:1471000} and the Future Circular Collider hadron--hadron option (FCC-hh)~\cite{Benedikt:2928193}, will operate with an extremely large number of simultaneous proton--proton interactions in each bunch crossing. Under such conditions, the association of charged-particle tracks with the correct hard-scattering vertex becomes increasingly difficult because signals from many interactions overlap in space and time (pile-up). To mitigate this pile-up contamination and maintain the physics performance of these experiments, precise timing information associated with charged-particle tracks is expected to play an important role, in addition to high spatial resolution~\cite{ATL-PHYS-PUB-2023-023}.

Low-Gain Avalanche Diodes (LGADs)~\cite{PELLEGRINI201412} are among the most promising sensor technologies for 4D tracking applications. LGADs are based on an $n^+$-in-$p$ structure with an additional highly doped $p^+$ gain layer located underneath the junction. This gain layer creates a strong local electric field, typically exceeding $300\,\mathrm{kV/cm}$, and induces controlled avalanche multiplication of signal carriers. Owing to this internal gain mechanism, LGADs can provide timing resolutions of $\mathcal{O}(10)\,\mathrm{ps}$, making them strong candidates for precision timing layers in future collider tracking detectors.

To further improve the spatial resolution while preserving the excellent timing performance, several extensions of the LGAD concept have been proposed. In particular, AC-coupled LGAD architectures, in which a continuous multiplication layer is combined with segmented AC readout, provide a promising route toward fine-pitch position-sensitive timing detectors with nearly full active area~\cite{nakamura2026developmentpixelatedcapacitivecoupledlgad,KITA2023168009}. The availability of detectors with both precise timing and fine spatial resolution in the inner tracker would significantly enhance the physics sensitivity of future collider experiments~\cite{ATL-PHYS-PUB-2023-023}.

For use in the innermost layers of future hadron collider experiments, LGADs require radiation tolerance up to fluences of the order of $10^{16}\,\neqcm$, where the performance of present devices degrades significantly. The main limitation is the radiation-induced degradation of the gain layer, commonly referred to as \textit{acceptor removal}~\cite{Moll:2020On}. In this process, electrically active acceptors in the gain layer are progressively deactivated under irradiation, resulting in a reduction of the effective acceptor concentration. This lowers the electric field in the multiplication region and hence reduces the internal gain of the sensor.

The loss of gain after irradiation can, in principle, be partially compensated by increasing the reverse bias voltage. For this reason, the increase in the operation voltage required to recover a given performance is a useful macroscopic measure of radiation damage in LGADs. Such compensation by higher bias is fundamentally limited by destructive breakdown phenomena such as single-event burnout (SEB)~\cite{tishelmancharny2026mortalityultrathinlgadspin,Beresford_2023}. Therefore, improving LGAD radiation tolerance requires not only operation at higher bias after irradiation, but also direct suppression of the gain-layer degradation itself.

A widely used phenomenological description of acceptor removal in a boron-only gain layer is the exponential decrease of the active acceptor concentration $N_{\mathrm{A}}$ as a function of irradiation fluence $\phi$~\cite{8331152},
\begin{equation}
    N_{\mathrm{A}}(\phi) = N_{\mathrm{A},0}\exp(-C_{\mathrm{A}}\phi),
    \label{eq:acceptor_removal}
\end{equation}
where $N_{\mathrm{A},0}$ is the initial active acceptor concentration and $C_{\mathrm{A}}$ is the acceptor-removal coefficient. A smaller value of $C_{\mathrm{A}}$ corresponds to slower degradation of the gain layer and hence to better radiation tolerance.

Several approaches have been proposed to mitigate acceptor removal and improve the radiation hardness of LGAD sensors, including modification of the oxygen~\cite{Imamura:2024zX} or carbon concentration in the gain layer~\cite{JIA2024169236} and the use of compensated gain-layer designs~\cite{SOLA2022167232}. In the latter approach, donors are intentionally introduced together with acceptors with the expectation that, if donor removal proceeds more rapidly than acceptor removal, the reduction of the effective acceptor excess over donors may be moderated after irradiation. Whether such an approach can improve the radiation tolerance of the gain layer is one of the questions addressed in this work.

In this work, we investigate HPK LGAD prototype sensors with modified gain-layer compositions, focusing on oxygen- and carbon-related modifications as well as donor-compensated structures. Their behavior is evaluated using the operation-voltage increase after irradiation together with the reduction of the effective acceptor concentration in the gain layer. The microscopic background of radiation effects relevant to LGAD gain layers is summarized in the next section. The experimental results are then presented, followed by a discussion of their physical interpretation.

\section{Radiation effects in LGAD gain layers}

Radiation damage in silicon detectors has been studied extensively for several decades, primarily in high-resistivity bulk material with doping concentrations of the order of $10^{12}\,\mathrm{cm^{-3}}$. In such lightly doped material, the macroscopic detector behavior is mainly governed by displacement-induced defects and their complexes, which modify the effective space charge, leakage current, and carrier trapping. By contrast, the implanted electrode regions of silicon detectors are typically doped above $\sim 10^{19}$--$10^{20}\,\mathrm{cm^{-3}}$, and are therefore only weakly affected by dopant deactivation because the concentration of displacement-induced defects remains far below the electrically active dopant density.

The LGAD gain layer lies between these two regimes. Its active acceptor concentration is typically of the order of $10^{16}\,\mathrm{cm^{-3}}$: much higher than that of the bulk, but not high enough for displacement-induced defect reactions to be negligible. For fluences relevant to the HL-LHC and beyond, the concentration of displacement-induced defects is typically estimated to be of the order of $10^{13}$--$10^{15}\,\mathrm{cm^{-3}}$, depending on the irradiation particle and the corresponding damage cross section. While this level is negligible compared with the dopant density in heavily doped electrodes, it is sufficiently large that a non-negligible fraction of the active acceptors in the LGAD gain layer can be deactivated through defect-mediated reactions. This intermediate doping regime is the key reason why LGADs are uniquely sensitive to acceptor removal, unlike either the low-doped bulk or the ultra-heavily doped electrode regions.

At the microscopic level, acceptor removal is generally described as a defect-mediated deactivation of boron atoms occupying regular silicon lattice sites (substitutional boron) in the gain layer. Irradiation creates primary displacement damage in the form of vacancies and silicon interstitials. The interstitial component can react with substitutional boron ($\mathrm{B_s}$), producing interstitial boron ($\mathrm{B_i}$), which is no longer an electrically active shallow acceptor in the same sense as $\mathrm{B_s}$. Subsequent reactions of $\mathrm{B_i}$ with impurities or other defects can form boron-related defect complexes. Among them, $\mathrm{B_iO_i}$, formed through a reaction with interstitial oxygen ($\mathrm{O_i}$), is one of the best-known examples, although it is not necessarily the only defect complex relevant to the evolution of the gain layer. These reactions reduce the concentration of electrically active acceptors in the gain layer and thereby lower the local electric field~\cite{8331152}.



Macroscopically, the resulting decrease of the active acceptor concentration is commonly parameterized by Eq.~(\ref{eq:acceptor_removal}). Although this description is phenomenological, the acceptor-removal coefficient $C_{\mathrm{A}}$ provides a practical and widely used figure of merit for comparing the radiation tolerance of different LGAD gain-layer designs and irradiation conditions.

The physical picture outlined above motivates several strategies for improving radiation tolerance. Carbon or oxygen modifications may alter the competition among defect reactions in the irradiated gain layer. Another possibility is a compensated gain-layer design, in which donor species are introduced together with acceptors. In such a structure, the effective acceptor excess over donors is determined by the difference between the active acceptor and donor concentrations,
\begin{equation}
    N_{\mathrm{eff}}(\phi)=N'_{\mathrm{A}}(\phi)-N_{\mathrm{D}}(\phi),
    \label{eq:neff_comp}
\end{equation}
where $N'_{\mathrm{A}}$ denotes the active acceptor concentration before subtraction of the donor contribution. The compensated structures are designed such that the initial effective concentration,
\begin{equation}
    N_{\mathrm{eff},0}=N'_{\mathrm{A},0}-N_{\mathrm{D},0},
    \label{eq:neff0_comp}
\end{equation}
is close to that of the boron-only reference design, so that similar multiplication characteristics are obtained before irradiation.

If both species are assumed to decrease exponentially under irradiation,
\begin{equation}
    N'_{\mathrm{A}}(\phi)=N'_{\mathrm{A},0}\exp(-C_{\mathrm{A}}\phi), \qquad
    N_{\mathrm{D}}(\phi)=N_{\mathrm{D},0}\exp(-C_{\mathrm{D}}\phi),
\end{equation}
then Eq.~(\ref{eq:neff_comp}) becomes
\begin{equation}
    N_{\mathrm{eff}}(\phi)=N'_{\mathrm{A},0}\exp(-C_{\mathrm{A}}\phi)-N_{\mathrm{D},0}\exp(-C_{\mathrm{D}}\phi).
    \label{eq:neff_comp_exp}
\end{equation}

In this simplified picture, compensation can improve radiation tolerance only if donor removal proceeds faster than acceptor removal, i.e. $C_{\mathrm{D}} > C_{\mathrm{A}}$. In that case, the donor contribution decreases more rapidly with fluence, and the reduction of the effective acceptor excess over donors may be moderated after irradiation. This expectation is based on a simplified picture in which acceptor and donor removal are treated as approximately independent processes.

In this work, these approaches are investigated experimentally through comparative measurements of irradiated prototype sensors. The results are presented in the following sections, and possible microscopic interpretations are discussed later.

\section{Samples and irradiation conditions}

\begin{figure}[tbp]
\centering
\includegraphics[width=0.3\linewidth]{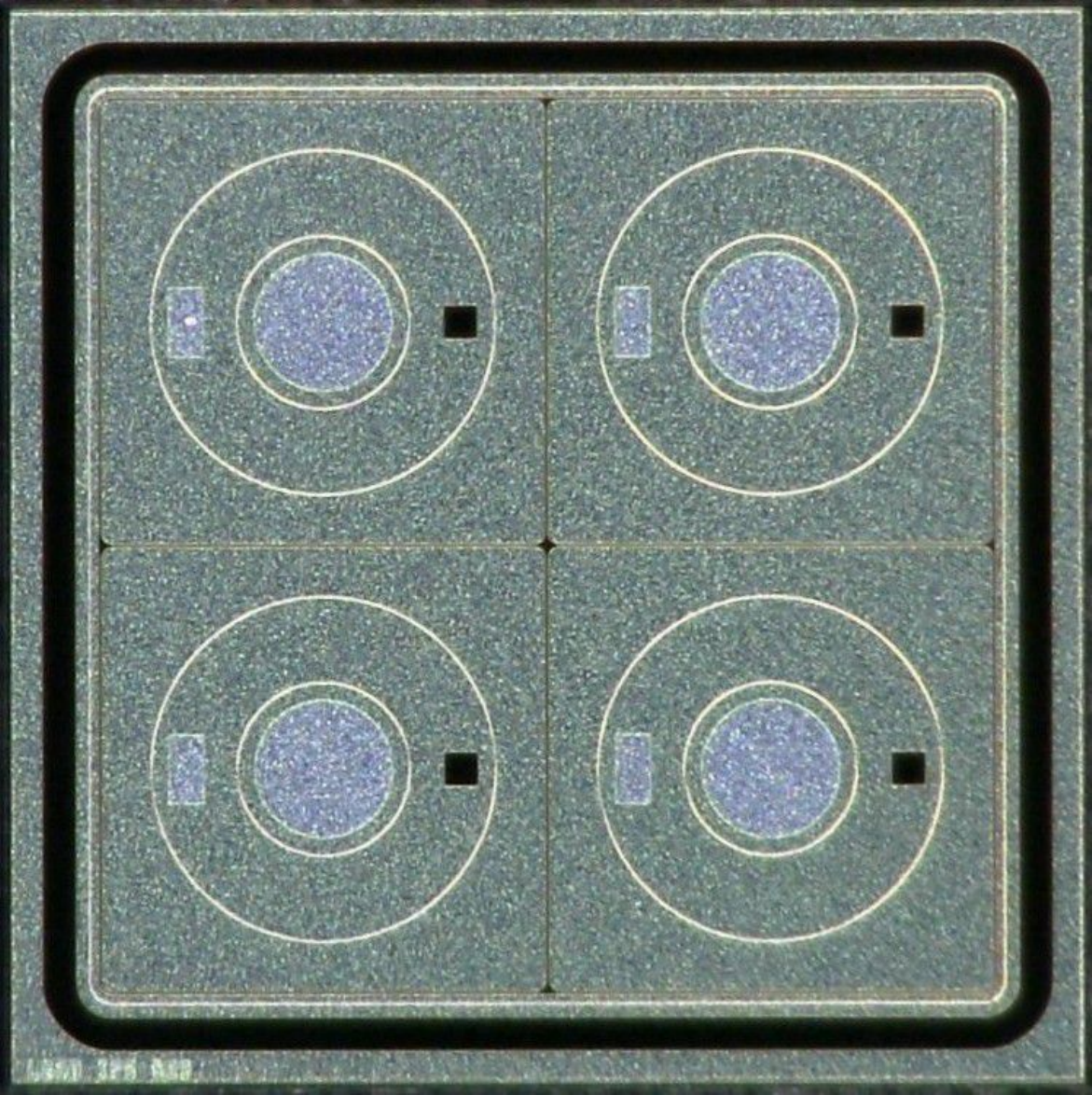}
\caption{Photograph of a representative HPK DC-LGAD prototype sensor used in this study. \label{fig:LGAD_picture}}
\end{figure}

To investigate possible approaches to mitigate radiation-induced degradation of the LGAD gain layer, several batches of prototype LGAD sensors were designed and fabricated in collaboration with Hamamatsu Photonics K.K. (HPK). An example of the evaluated prototype is shown in Fig.~\ref{fig:LGAD_picture}. All sensors used in this study are DC-LGADs consisting of an $n^{++}$ readout electrode and a $p^+$ gain layer on a $p$-type substrate. The active thickness is $50\,\um{}$, and the sensor is segmented into a $2\times 2$ pixel array with a pitch of $1.3\,\mathrm{mm}$.

Three design approaches were investigated in this work: oxygen-related modification, carbon implantation, and gain-layer compensation.

The first approach concerns oxygen-related modification of the gain layer. In our earlier HPK productions, a long high-temperature annealing process was applied to stabilize the boron activation. As a consequence, oxygen contamination diffusing from the substrate into the epitaxial layer became much larger than the boron concentration in the gain layer. In the present production, the annealing time was reduced to the minimum required for stable processing, while keeping the same diffusion path from the substrate to the epitaxial layer. This process optimization reduced the oxygen contamination by more than one order of magnitude.
In addition to a high-oxygen sample (B+O) produced with the conventional thermal process, a low-oxygen sample (B), used here as the reference sample, was produced with the shortened thermal process. Furthermore, a Partially Activated Boron (PAB) variant was fabricated in the reference production to study the effect of boron--oxygen related defects such as $\mathrm{B_iO_i}$~\cite{Imamura:2024zX}. In this sample, additional inactive boron was introduced as a sacrificial component intended to clean oxygen-related reactions before electrically active boron in the gain layer is affected.

The second approach is carbon implantation into the gain layer. In these samples, carbon was co-doped at the wafer level on the sensor surface before the device processing, and was subsequently diffused over the gain-layer region during the thermal process. To evaluate the effect of carbon on radiation tolerance in the present HPK production, carbon-implanted samples were prepared for both the low-oxygen and high-oxygen process conditions, denoted as B+C and B+C+O, respectively.

The third approach is gain-layer compensation using co-implantation of boron and phosphorus. The key design requirement for these samples was to match the depth profiles of boron and phosphorus appropriately so that the compensated structures retained multiplication characteristics close to those of the boron-only reference design. By tuning the implantation parameters, compensated samples were produced with gain and breakdown voltage comparable to the standard boron-only structures. In the present prototypes, compensated structures were prepared for both low-oxygen and high-oxygen conditions, denoted as 2B+1P and 2B+1P+O, respectively. In addition, a compensated sample with carbon implantation, 2B+1P+C, was included in order to study the combined effect of compensation and carbon.

The full set of prototype samples evaluated in this study is summarized in Table~\ref{tab:samples}. The table includes the reference sample, the high-oxygen sample, and the carbon-implanted and compensated variants.

\begin{table}[htbp]
\centering
\caption{List of prototype LGAD samples tested in this study.}
\label{tab:samples}
\begin{tabular}{lccc}
\hline
Sample & Oxygen condition & Carbon & Compensation \\
\hline
B (reference)   & Low  & No  & No \\
B+O            & High & No  & No \\
PAB            & Low  & No  & No \\
B+C            & Low  & Yes & No \\
B+C+O          & High & Yes & No \\
2B+1P          & Low  & No  & Yes \\
2B+1P+O        & High & No  & Yes \\
2B+1P+C        & Low  & Yes & Yes \\
\hline
\end{tabular}
\end{table}

\subsection{Irradiation campaigns}

The radiation tolerance of the prototype sensors was evaluated through both proton and neutron irradiation campaigns. Proton irradiations were carried out at the Research Center for Accelerator and Radioisotope Science (RARiS, formerly CYRIC), Tohoku University, Japan, while neutron irradiations were performed at the TRIGA Mark II reactor of the Jo\v{z}ef Stefan Institute (JSI), Slovenia~\cite{refId0}.

The proton beam at RARiS was provided by the AVF cyclotron. Two proton irradiation campaigns were carried out in 2025 with beam energies of $45\,\mathrm{MeV}$ and $70\,\mathrm{MeV}$, respectively. During irradiation, the samples were cooled to approximately $-15\,^\circ\mathrm{C}$ using a liquid-nitrogen-based cooling system in order to suppress annealing during exposure. Immediately after irradiation, the samples were stored in a freezer until electrical characterization~\cite{Nakamura_2015}.

For proton irradiation, the accumulated fluence for each irradiation stack was determined from the activation of aluminum foils attached to the front or back side of the sample holder. The fluence was estimated from the measured $\gamma$-ray activity of ${}^{24}\mathrm{Na}$ produced in the aluminum during irradiation. For neutron irradiation, the fluence values were taken from the standard JSI TRIGA reactor dosimetry and quoted in terms of the corresponding $1\,\mathrm{MeV}$ neutron-equivalent fluence~\cite{refId0}. A systematic uncertainty of 10\% was assigned to both the proton and neutron fluence values used in this work.

The first proton campaign was carried out in January 2025 using a $45\,\mathrm{MeV}$ proton beam. The beam energy was lower than the standard RARiS operating condition because of a temporary limitation in the Dee electrode voltage of the cyclotron. This campaign was used primarily to evaluate the oxygen-modified samples, namely B, B+O, and PAB. In addition, the dataset provides information on the proton-energy dependence of radiation damage when compared with the $70\,\mathrm{MeV}$ irradiation.

The second proton campaign was performed in November 2025 using a $70\,\mathrm{MeV}$ proton beam. This irradiation was used to evaluate the carbon-implanted and compensated gain-layer designs. The sample set included the reference sample B and the high-oxygen sample B+O, the carbon-implanted samples B+C and B+C+O, and the compensated structures 2B+1P, 2B+1P+O, and 2B+1P+C.

To compare the gain-layer degradation under different irradiation particles, a subset of the prototype sensors was also irradiated with reactor neutrons at the JSI TRIGA Mark II reactor. The neutron irradiation was used mainly to compare the acceptor-removal behavior with the proton data for the same sample set. After neutron irradiation, the samples were stored in a freezer until characterization.

Throughout this paper, irradiation fluences are quoted in terms of the $1\,\mathrm{MeV}$ neutron-equivalent fluence, $\Phi_{\mathrm{eq}}$, based on the Non-Ionizing Energy Loss (NIEL) scaling hypothesis. For proton irradiation, the hardness factors used for the conversion were $\kappa = 1.91$ for $45\,\mathrm{MeV}$ protons and $\kappa = 1.53$ for $70\,\mathrm{MeV}$ protons~\cite{NIEL}.

\section{Measurement setup and analysis method}

To evaluate the radiation tolerance of the prototype sensors, electrical characterizations were performed before and after irradiation. All measurements described below were carried out in a temperature-controlled environment at approximately $-20\,^\circ\mathrm{C}$. Low-temperature operation is essential in order to suppress the leakage current enhanced by radiation-induced defects and to minimize further annealing during the measurements.

\subsection{Measurement setup}

\subsubsection{Current--voltage measurement}

Current--voltage (IV) measurements were performed to evaluate the electrical behavior of the irradiated sensors and, in particular, the evolution of the gain-layer depletion voltage. A Keithley 2410 source measurement unit was used to apply the reverse bias voltage and to record the leakage current.

\subsubsection{Beta-ray timing measurement}

\begin{figure}[tbp]
\centering
\includegraphics[width=\linewidth]{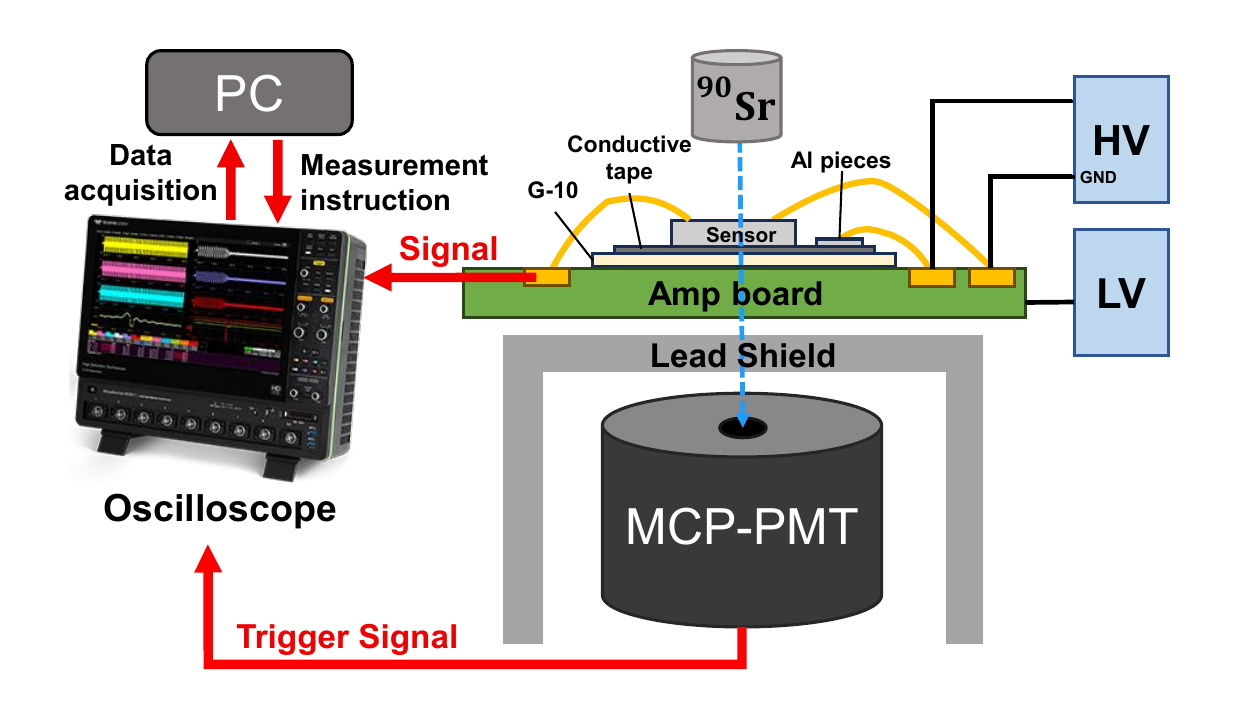}
\caption{Schematic view of the ${}^{90}\mathrm{Sr}$ beta-ray timing measurement setup used to evaluate the pulse-height and timing performance of the LGAD samples. \label{fig:betaray_setup}}\end{figure}

To investigate the timing performance of the irradiated samples and its dependence on the bias voltage, measurements were performed using a ${}^{90}\mathrm{Sr}$ beta-ray source. The setup employed a coincidence trigger with a micro-channel-plate photomultiplier tube (MCP-PMT, Hamamatsu R3809U-52) as the timing reference. Its time resolution is sufficiently small compared with that of the LGAD samples studied here, and its contribution to the measured timing resolution is therefore neglected.

The LGAD sensors were mounted on a 16-channel amplifier board developed at KEK. The board employs a two-stage discrete RF amplifier configuration using Mini-Circuits GALI-S66+ amplifiers in order to provide high-bandwidth signal readout. The waveforms were digitized with a high-speed oscilloscope (LeCroy WaveRunner 8208HD) with a bandwidth of $2\,\mathrm{GHz}$ and a sampling rate of $10\,\mathrm{GS/s}$.

\subsection{Analysis observables}

\subsubsection{Gain-layer depletion voltage and acceptor-removal coefficient}

To evaluate the acceptor-removal effect through a macroscopic electrical observable, the gain-layer depletion voltage, $V_{\mathrm{gl}}$, was extracted from the IV curves. Under reverse bias, the depletion width of the sensor, denoted by $W$, increases with the applied voltage. The dark current of an irradiated sensor is dominated by the generation current in the depleted region, $I_{\mathrm{gen}}$, which scales approximately with $W$. In this approximation, the measured dark current $I$ may be written as $I \approx I_{\mathrm{gen}} \propto W$. Therefore, the logarithmic derivative of the current,
\begin{equation}
    \frac{d(\ln I)}{dV} \approx \frac{1}{W}\frac{dW}{dV},
\end{equation}
enhances changes in the depletion behavior. When the gain layer becomes depleted, the voltage dependence of $W$ changes, producing a characteristic feature in the IV curve. In this work, $V_{\mathrm{gl}}$ was operationally defined as the voltage at which the logarithmic derivative of the current, $d(\ln I)/dV$, reaches its maximum.

For samples with the same geometry, the gain-layer depletion voltage is approximately proportional to the active acceptor concentration in the gain layer, $V_{\mathrm{gl}} \propto N_{\mathrm{A}}$. Its fluence dependence can therefore be parameterized as
\begin{equation}
V_{\mathrm{gl}}(\phi) = V_{\mathrm{gl},0}\exp(-C_{\mathrm{A}}\phi),
\label{eq:Vgl}
\end{equation}
where $V_{\mathrm{gl},0}$ is the gain-layer depletion voltage before irradiation and $C_{\mathrm{A}}$ is the acceptor-removal coefficient. By fitting the measured $V_{\mathrm{gl}}$ values as a function of fluence, $C_{\mathrm{A}}$ was determined for each sample type.

\subsubsection{Definition of the operation voltage}

The timing performance was evaluated from the waveforms obtained in the beta-ray measurements. The signal pulse height was defined as the most probable value (MPV) of the Landau distribution fitted to the pulse-height spectrum. The timing resolution, $\sigma_t$, was extracted from the distribution of the time difference between the LGAD signal and the MCP-PMT reference signal using a constant-fraction method with a threshold fraction of 50\%.

For each sample, the timing resolution was measured as a function of the bias voltage. The operation voltage, $V_{\mathrm{op}}$, was then defined as the bias voltage at which the best timing resolution was achieved within the measured range. This quantity provides a practical indicator of the voltage scale associated with recovering the highest timing performance after irradiation. In practice, however, detector operation does not necessarily require biasing the sensor exactly at $V_{\mathrm{op}}$, since the timing requirement may already be satisfied at a somewhat lower voltage. In this sense, $V_{\mathrm{op}}$ is used here primarily as a comparative figure of merit for radiation tolerance.

\section{Experimental results}
\label{sec:results}

In this section, the radiation tolerance of the prototype sensors is presented using two complementary observables: the acceptor-removal coefficient $C_{\mathrm{A}}$, extracted from IV measurements through the fluence dependence of the gain-layer depletion voltage $V_{\mathrm{gl}}$, and the operation voltage $V_{\mathrm{op}}$, determined from the beta-ray timing measurements.

\subsection{Pre-irradiation electrical and timing characteristics}

\begin{figure}[tbp]
\centering
\begin{subfigure}[t]{0.48\linewidth}
  \centering
  \includegraphics[width=\linewidth]{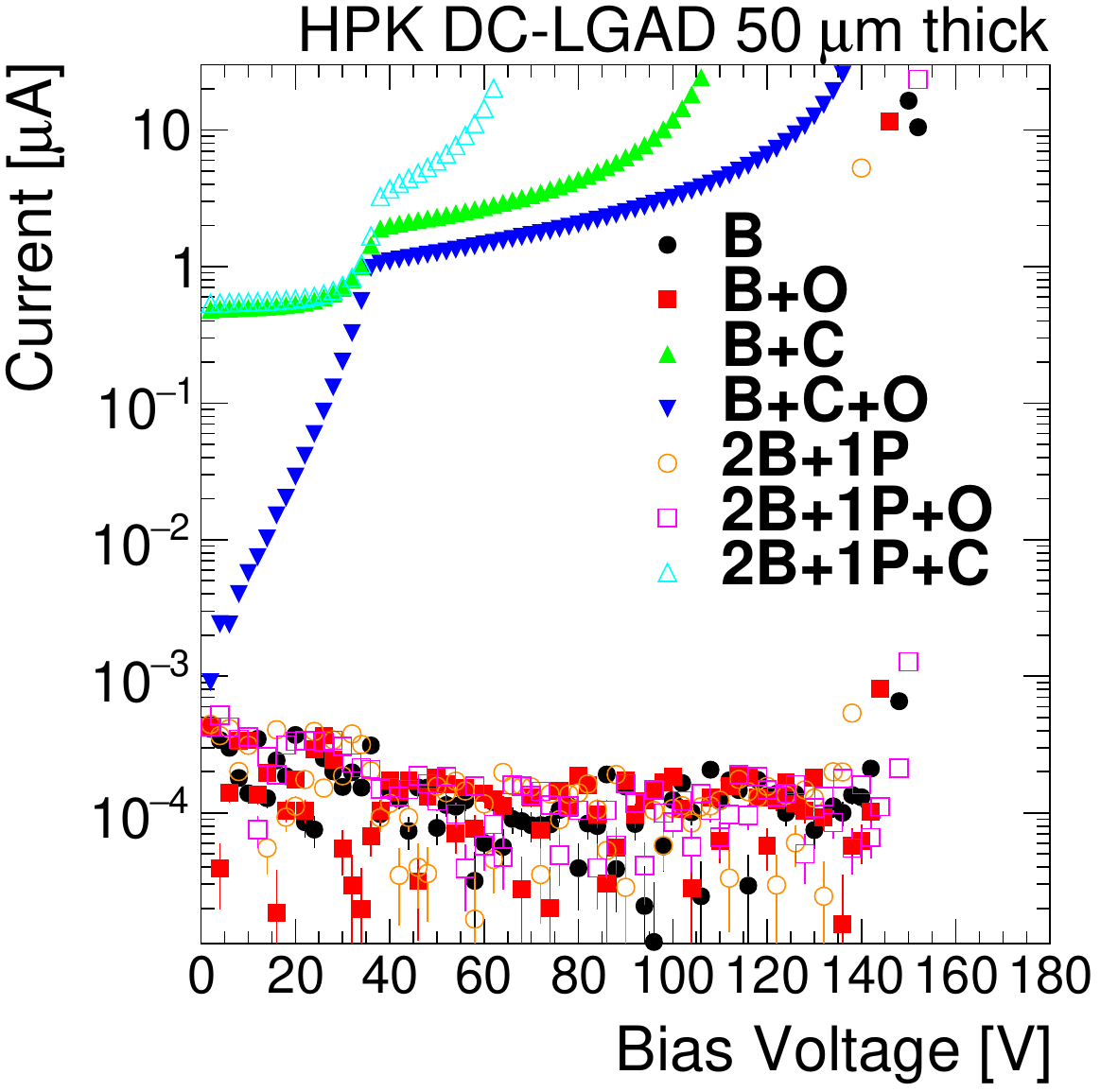}
  \caption{IV characteristics}
\end{subfigure}
\hfill
\begin{subfigure}[t]{0.48\linewidth}
  \centering
  \includegraphics[width=\linewidth]{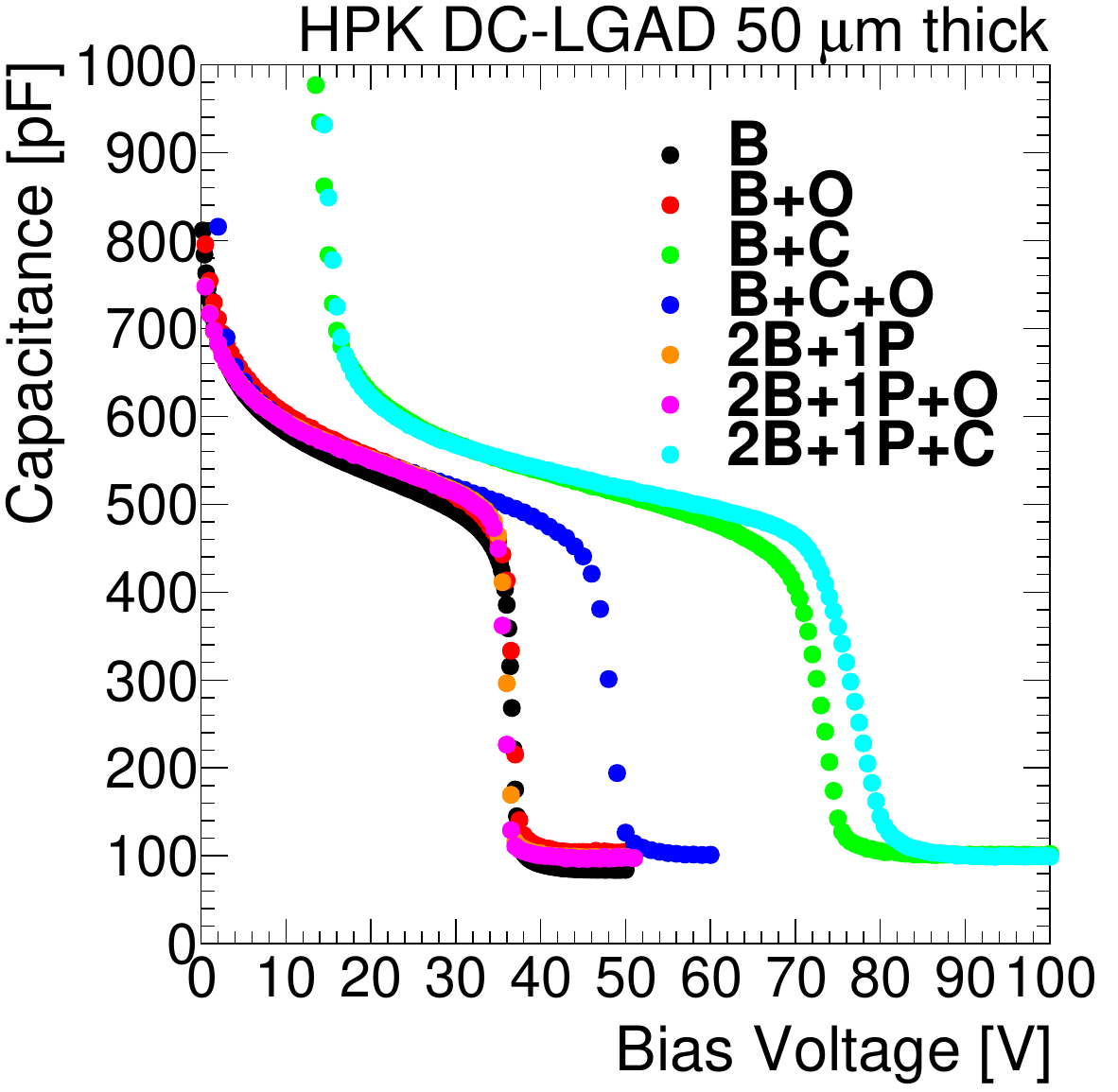}
  \caption{CV characteristics}
\end{subfigure}

\begin{subfigure}[t]{0.48\linewidth}
  \centering
  \includegraphics[width=\linewidth]{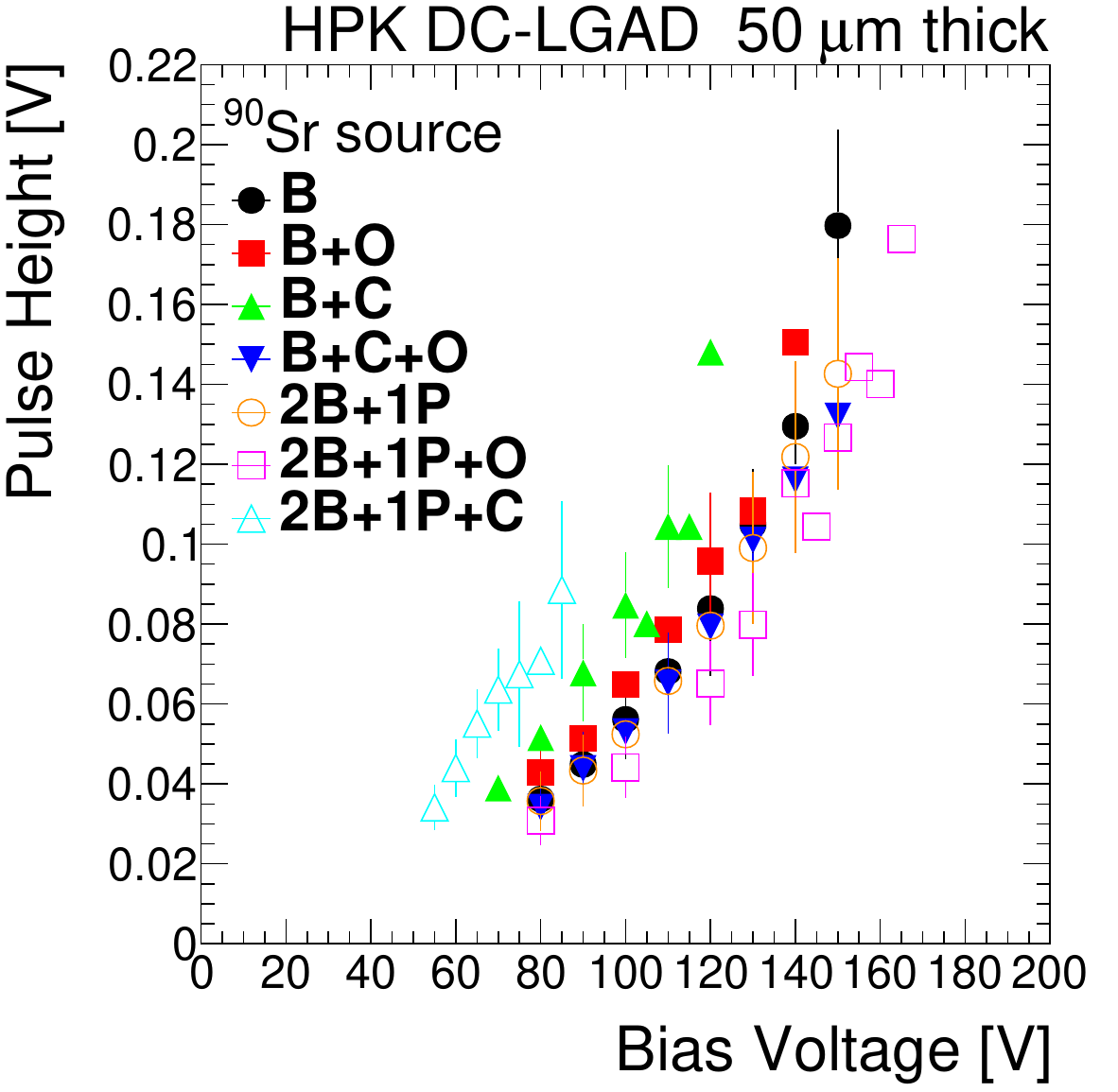}
  \caption{Pulse height}
\end{subfigure}
\hfill
\begin{subfigure}[t]{0.48\linewidth}
  \centering
  \includegraphics[width=\linewidth]{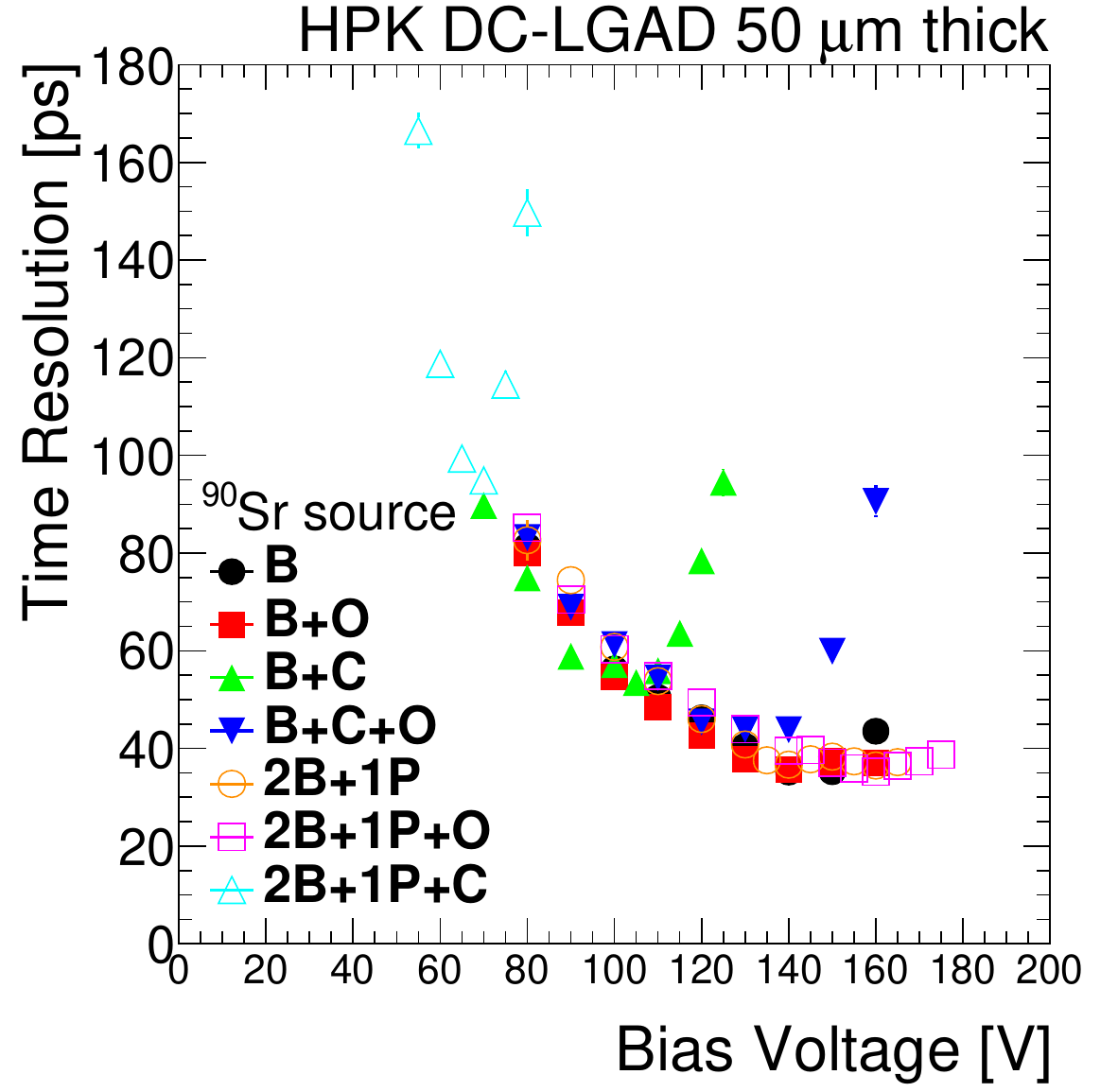}
  \caption{Time resolution}
\end{subfigure}
\caption{Pre-irradiation electrical and timing characteristics of the reference, high-oxygen, carbon-implanted, and compensated samples as a function of bias voltage: (a) IV characteristics, (b) CV characteristics, (c) pulse height, and (d) timing resolution. The carbon-implanted samples show increased leakage current and earlier breakdown, while their pulse-height and timing behavior remain comparable to those of the other samples over most of the operating range.}
\label{fig:pre_rad_iv_timing}
\end{figure}

Before discussing the irradiated samples, the electrical and timing characteristics of the non-irradiated carbon-implanted and compensated structures were evaluated. Figure~\ref{fig:pre_rad_iv_timing} shows the bias-voltage dependence of the IV characteristics, CV characteristics, pulse height, and timing resolution before irradiation. In the CV curves, $V_{\mathrm{gl}}$ approximately corresponds to the voltage region where the capacitance starts to decrease rapidly after the relatively flat plateau at high capacitance.

As shown in Fig.~\ref{fig:pre_rad_iv_timing}(a), the non-carbon samples exhibit similar IV characteristics and comparable breakdown voltages before irradiation. In contrast, the carbon-implanted samples show substantially larger leakage currents and earlier breakdown. The CV characteristics in Fig.~\ref{fig:pre_rad_iv_timing}(b) indicate that the gain-layer depletion voltage is well controlled and consistent among the non-carbon samples, while the carbon-implanted samples exhibit slightly higher $V_{\mathrm{gl}}$ values. Therefore, the earlier breakdown observed in the carbon-implanted structures is not associated with a reduced gain-layer depletion voltage.

The pulse-height curves shown in Fig.~\ref{fig:pre_rad_iv_timing}(c) are in good agreement among all samples over the measured voltage range, indicating that the signal amplitude before irradiation is similar for the different gain-layer designs. The timing-resolution curves in Fig.~\ref{fig:pre_rad_iv_timing}(d) also show the same basic voltage dependence. However, for the carbon-implanted samples, the timing resolution degrades rapidly in the voltage region close to breakdown, consistent with the increased noise associated with their larger leakage current. These pre-irradiation characteristics show that the carbon-implanted samples differ mainly in leakage current and breakdown behavior, while their initial gain-layer depletion voltage is not reduced relative to the other structures. Therefore, the post-irradiation results presented below can be used directly to assess the relative radiation tolerance of the different gain-layer designs.

\subsection{Gain-layer degradation from IV and CV measurements}

The gain-layer degradation after irradiation was evaluated through the fluence dependence of the gain-layer depletion voltage $V_{\mathrm{gl}}$ extracted from the IV measurements. The corresponding acceptor-removal coefficients $C_{\mathrm{A}}$ were obtained by fitting the measured $V_{\mathrm{gl}}$ values with Eq.~\eqref{eq:Vgl}.

\begin{figure}[tbp]
\centering
\includegraphics[width=0.6\linewidth]{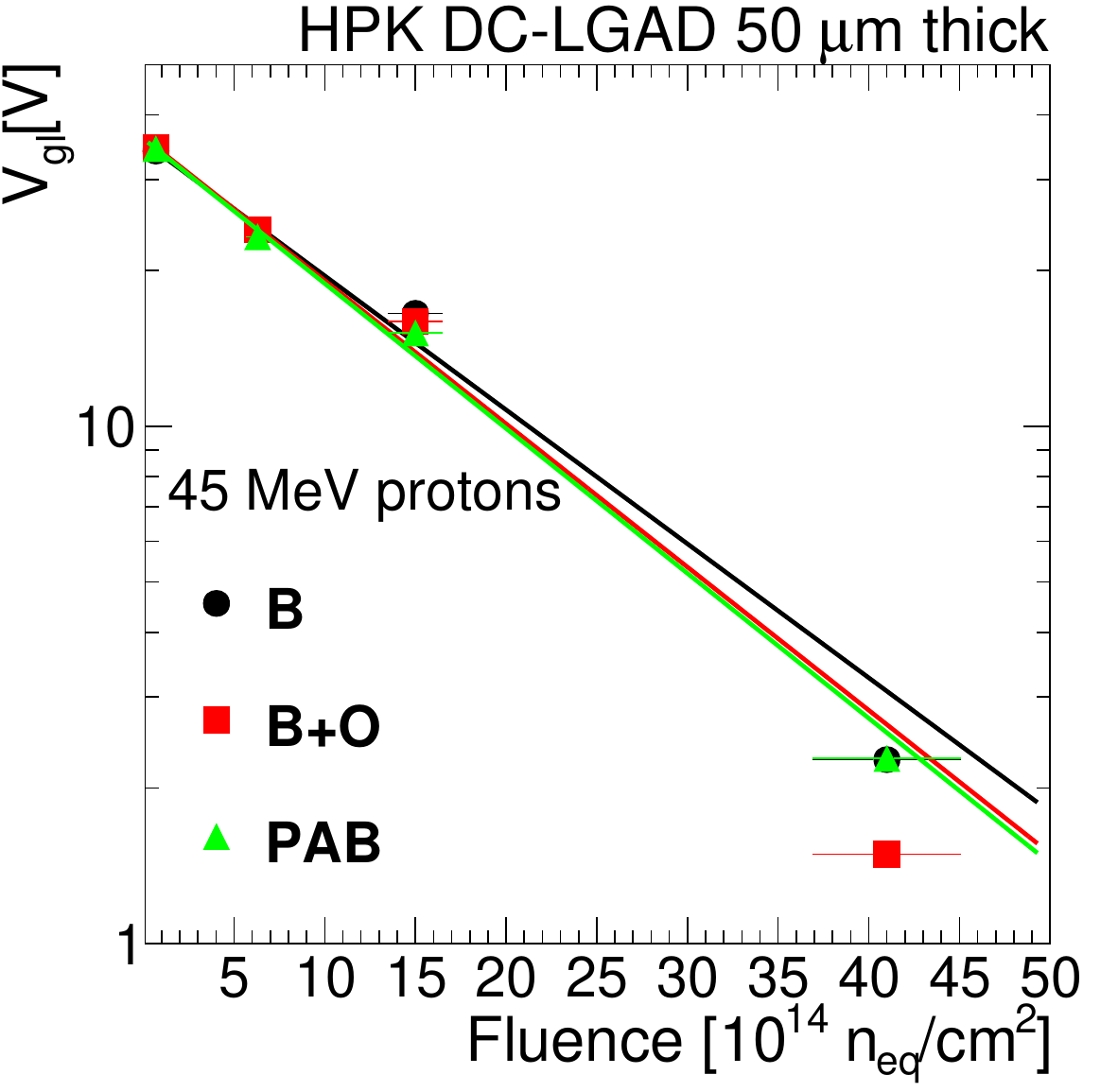}
\caption{Fluence dependence of the gain-layer depletion voltage $V_{\mathrm{gl}}$ for the oxygen-modified samples irradiated with $45\,\mathrm{MeV}$ protons. The curves show fits with Eq.~\eqref{eq:Vgl} used to extract the acceptor-removal coefficient $C_{\mathrm{A}}$. \label{fig:Vgl_oxygen}}
\end{figure}

The oxygen-modified samples and Partially Activated Boron (PAB) sample were irradiated with $45\,\mathrm{MeV}$ protons up to a fluence of $4.1\times10^{15}\,\neqcm$. Prior to the measurements, these samples were annealed at $60^\circ\mathrm{C}$ for 80 minutes. Figure~\ref{fig:Vgl_oxygen} shows the fluence dependence of $V_{\mathrm{gl}}$ for the oxygen-modified samples, and the extracted $C_{\mathrm{A}}$ values are summarized in Table~\ref{tab:cA_oxygen}. Within the present uncertainties, the fitted $C_{\mathrm{A}}$ values are consistent among the B, B+O, and PAB samples.

\begin{table}[htbp]
\centering
\caption{Acceptor-removal coefficients $C_A$ for oxygen-modified samples irradiated with $45\,\mathrm{MeV}$ protons.}
\label{tab:cA_oxygen}
\begin{tabular}{lc}
\hline
Sample & $C_A$ [$10^{-16}\,\mathrm{n_{eq}\,cm}^2$] \\
\hline
B   & $5.96 \pm 0.39$ \\
B+O & $6.37 \pm 0.44$ \\
PAB & $6.45 \pm 0.40$ \\
\hline
\end{tabular}
\end{table}

\begin{figure}[tbp]
\centering
\includegraphics[width=0.6\linewidth]{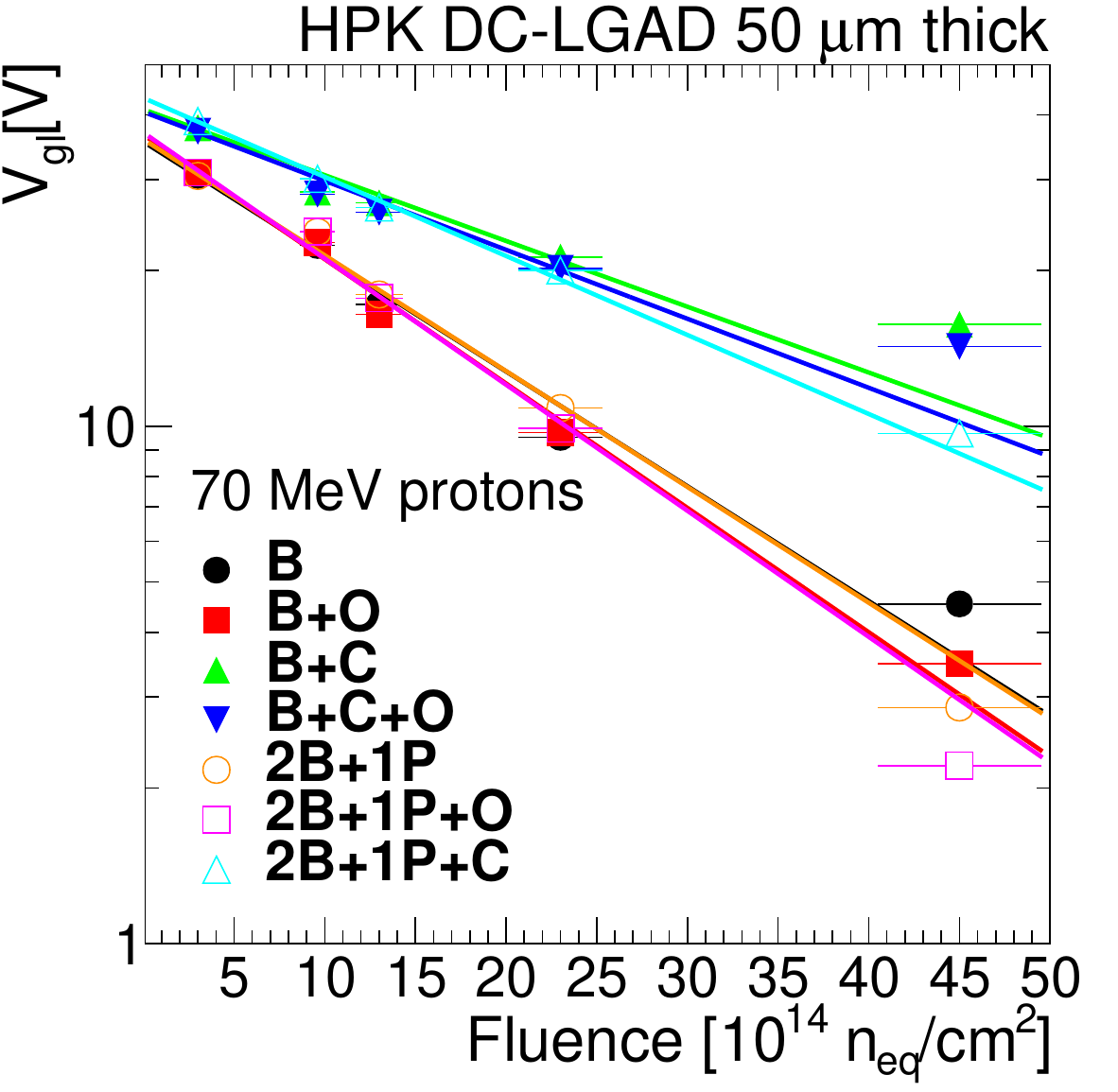}
\caption{Fluence dependence of the gain-layer depletion voltage $V_{\mathrm{gl}}$ for the reference and high-oxygen boron-only, carbon-implanted, and compensated samples irradiated with $70\,\mathrm{MeV}$ protons. The curves show fits with Eq.~\eqref{eq:Vgl} used to extract the acceptor-removal coefficient $C_{\mathrm{A}}$. \label{fig:Vgl_CarbonComp}}
\end{figure}

The carbon-implanted and compensated samples were evaluated using $70\,\mathrm{MeV}$ proton irradiation up to $4.5\times10^{15}\,\neqcm$. In contrast to the oxygen-modified sample study, these measurements were performed without an additional annealing step after irradiation. Figure~\ref{fig:Vgl_CarbonComp} shows the fluence dependence of $V_{\mathrm{gl}}$, and the extracted $C_{\mathrm{A}}$ values are summarized in Table~\ref{tab:cA_comp}.

\begin{table}[htbp]
\centering
\caption{Acceptor-removal coefficients $C_A$ for the carbon-implanted and compensated samples irradiated with $70\,\mathrm{MeV}$ protons.}
\label{tab:cA_comp}
\begin{tabular}{lc}
\hline
Sample & $C_A$ [$10^{-16}\,\mathrm{n_{eq}\,cm}^2$] \\
\hline
B         & $5.11 \pm 0.30$ \\
B+O       & $5.53 \pm 0.34$ \\
B+C       & $2.93 \pm 0.16$ \\
B+C+O     & $3.07 \pm 0.17$ \\
2B+1P     & $5.15 \pm 0.36$ \\
2B+1P+O   & $5.61 \pm 0.41$ \\
2B+1P+C   & $3.52 \pm 0.22$ \\
\hline
\end{tabular}
\end{table}

The reference sample B and the high-oxygen sample B+O exhibit consistent $C_{\mathrm{A}}$ values within the present experimental uncertainties. The compensation-only samples 2B+1P and 2B+1P+O also exhibit $C_{\mathrm{A}}$ values similar to those of the corresponding non-carbon reference samples. By contrast, the carbon-implanted samples B+C and B+C+O show substantially smaller $C_{\mathrm{A}}$ values. The combined sample 2B+1P+C also shows a reduced $C_{\mathrm{A}}$ compared with the non-carbon compensated structures, but its fitted value remains slightly larger than that of the carbon-only sample B+C.

\subsection{Timing performance after irradiation}
\begin{figure}[tbp]
\centering
\begin{subfigure}[t]{0.48\linewidth}
  \centering
  \includegraphics[width=\linewidth]{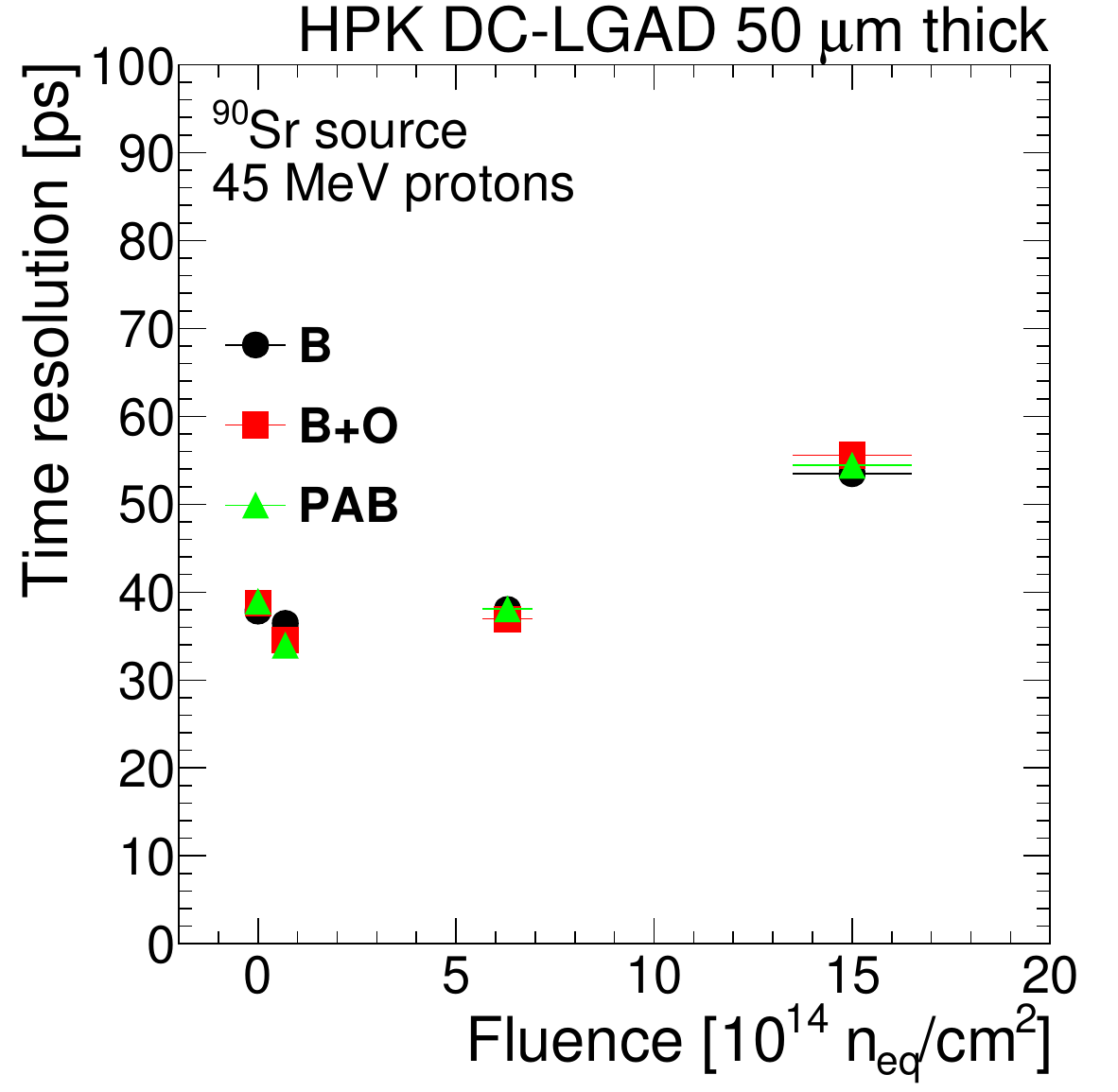}
  \caption{Time resolution}
\end{subfigure}
\hfill
\begin{subfigure}[t]{0.48\linewidth}
  \centering
  \includegraphics[width=\linewidth]{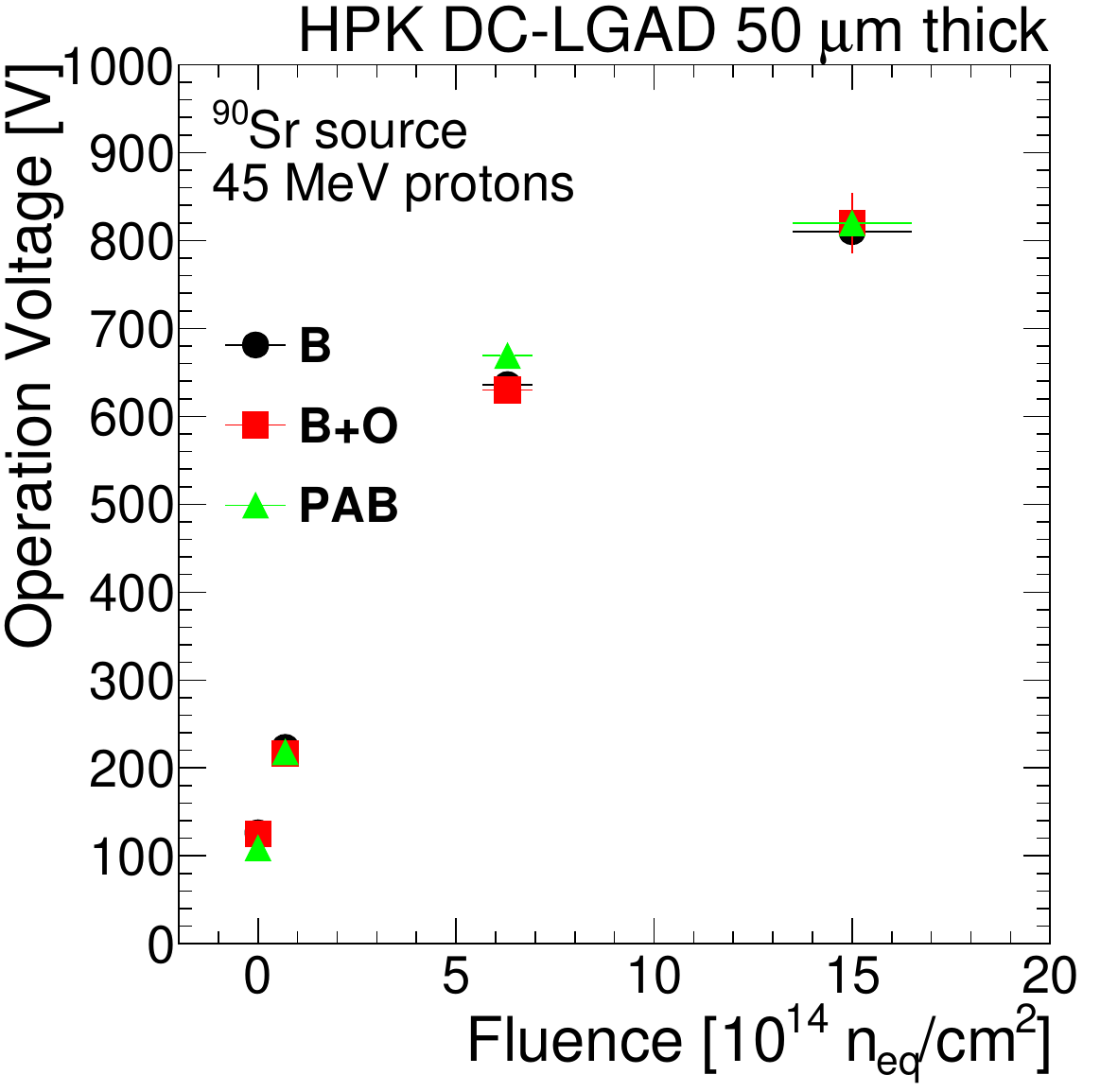}
  \caption{Operation voltage}
\end{subfigure}
\caption{Fluence dependence of the timing performance for the oxygen-modified samples irradiated with $45\,\mathrm{MeV}$ protons: (a) timing resolution and (b) operation voltage $V_{\mathrm{op}}$. No clear improvement is observed among B, B+O, and PAB within the present uncertainties.\label{fig:treso_vs_fluence_oxygen}}
\end{figure}

The timing performance after irradiation was evaluated using the beta-ray measurements. For each sample, the timing resolution was measured as a function of the applied bias voltage, and the operation voltage $V_{\mathrm{op}}$ was defined as described in Section~4. Figure~\ref{fig:treso_vs_fluence_oxygen} shows the fluence dependence of the timing resolution and $V_{\mathrm{op}}$ for the oxygen-modified samples irradiated with $45\,\mathrm{MeV}$ protons.

At the highest fluence point of $4.1\times10^{15}\,\neqcm$, the timing signal of the oxygen-modified samples could no longer be reliably separated from the noise, and this point was therefore excluded from the timing analysis. Over the remaining fluence range, no clear separation is observed among the B, B+O, and PAB samples in either the timing resolution or $V_{\mathrm{op}}$.

\begin{figure}[tbp]
\centering
\begin{subfigure}[t]{0.48\linewidth}
  \centering
  \includegraphics[width=\linewidth]{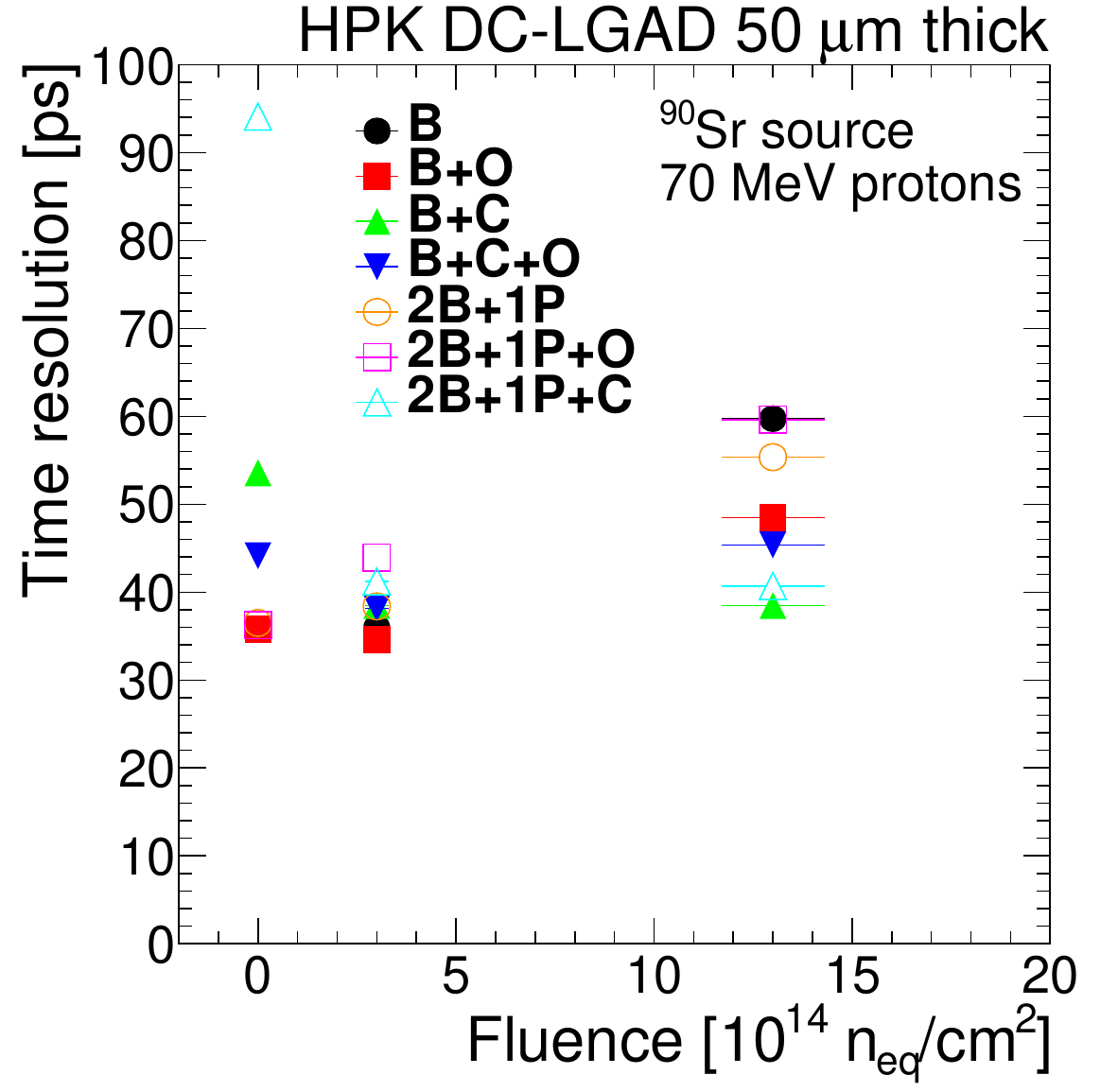}
  \caption{Time resolution}
\end{subfigure}
\hfill
\begin{subfigure}[t]{0.48\linewidth}
  \centering
  \includegraphics[width=\linewidth]{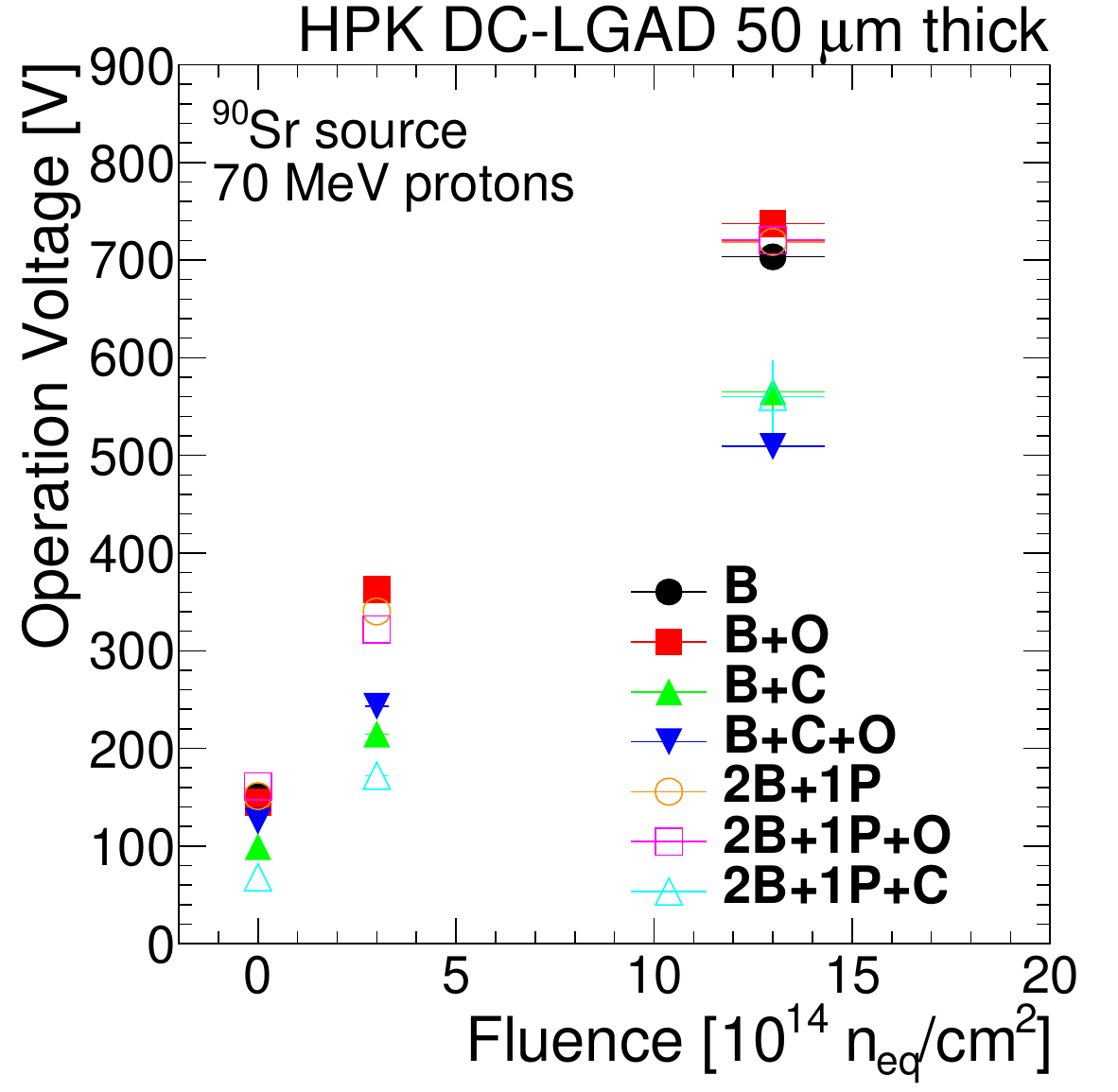}
  \caption{Operation voltage}
\end{subfigure}
\caption{Fluence dependence of the timing performance for the reference and high-oxygen boron-only, carbon-implanted, and compensated samples irradiated with $70\,\mathrm{MeV}$ protons: (a) timing resolution and (b) operation voltage $V_{\mathrm{op}}$. The carbon-implanted samples require lower bias voltage after irradiation while maintaining timing performance comparable to that of the other sample groups.}
\label{fig:treso_vs_fluence_comp}
\end{figure}

The corresponding timing results for the oxygen-modified, carbon-implanted, and compensated samples irradiated with $70\,\mathrm{MeV}$ protons are shown in Fig.~\ref{fig:treso_vs_fluence_comp}. Although the gain layer degrades with irradiation, the timing resolution remains nearly unchanged up to a fluence of  $1.3\times10^{15}\,\neqcm$ when the applied bias voltage is increased to compensate for the reduction of the gain-layer electric field. The oxygen-modified samples show no clear difference in $V_{\mathrm{op}}$ compared with the corresponding reference sample, consistent with the observation in the $45\,\mathrm{MeV}$ dataset. Likewise, the compensation-only sample 2B+1P exhibits a fluence dependence of the timing resolution and $V_{\mathrm{op}}$ similar to that of the reference sample B, and the oxygen-modified compensated sample 2B+1P+O also behaves similarly to 2B+1P within the present measurement precision.

As discussed in the pre-irradiation characterization, the carbon-implanted samples show somewhat degraded timing performance before irradiation, which also leads to an underestimated initial $V_{\mathrm{op}}$. After irradiation, however, the carbon-implanted samples B+C and B+C+O exhibit clearly lower operation voltages than the corresponding non-carbon samples. The combined sample 2B+1P+C also shows timing performance comparable to that of the other carbon-implanted samples and clearly better than that of the non-carbon compensated samples. Within the present dataset, the post-irradiation timing results therefore show a clear improvement only for the carbon-implanted structures.

\begin{figure}[tbp]
\centering
\includegraphics[width=0.68\linewidth]{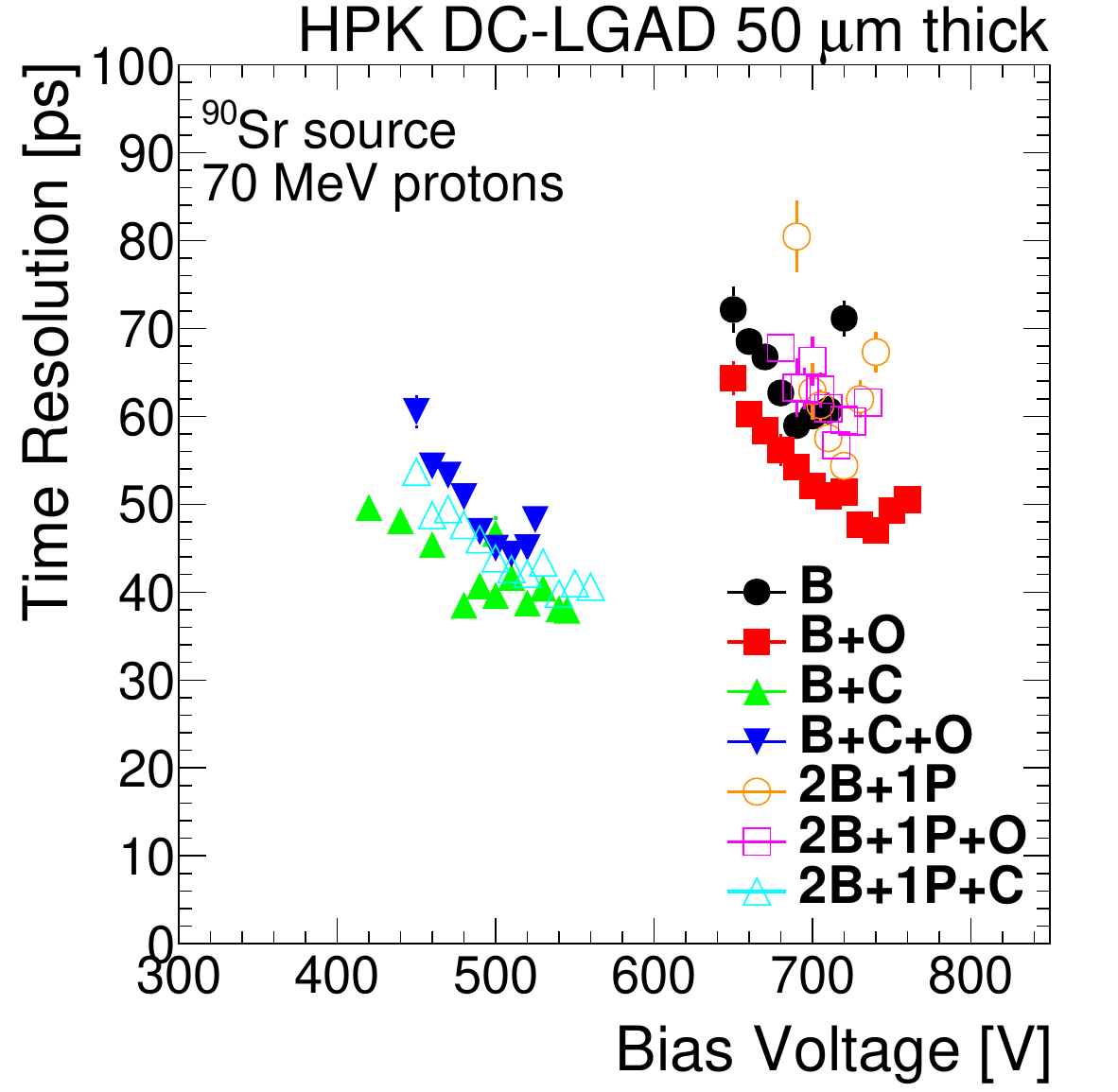}
\caption{Timing resolution as a function of bias voltage at a fluence of $1.3\times10^{15}\,\neqcm$ for the reference, oxygen-modified, carbon-implanted, and compensated samples. The carbon-implanted samples are shifted toward lower operating voltage while achieving timing resolution comparable to that of the other sample groups.}
\label{fig:tres_vs_bias_1p3e15}
\end{figure}

To visualize the difference among the gain-layer designs more directly, Fig.~\ref{fig:tres_vs_bias_1p3e15} compares the timing resolution as a function of the bias voltage at a fluence of $1.3\times10^{15}\,\neqcm$. The oxygen-modified samples and the compensation-only samples populate the same high-voltage region as the reference sample, indicating no visible reduction of the voltage required to recover the timing performance. By contrast, the carbon-implanted samples are clearly shifted toward lower bias voltage while reaching a similar level of timing resolution. The combined sample 2B+1P+C is located in the same low-voltage region as the other carbon-implanted samples. This representation shows directly that, among the methods studied here, only carbon implantation leads to a clear reduction of the bias voltage required to maintain the timing performance after irradiation.

\subsection{Correlation between $C_{\mathrm{A}}$ and post-irradiation operation voltage}

\begin{figure}[tbp]
\centering
\includegraphics[width=0.65\linewidth]{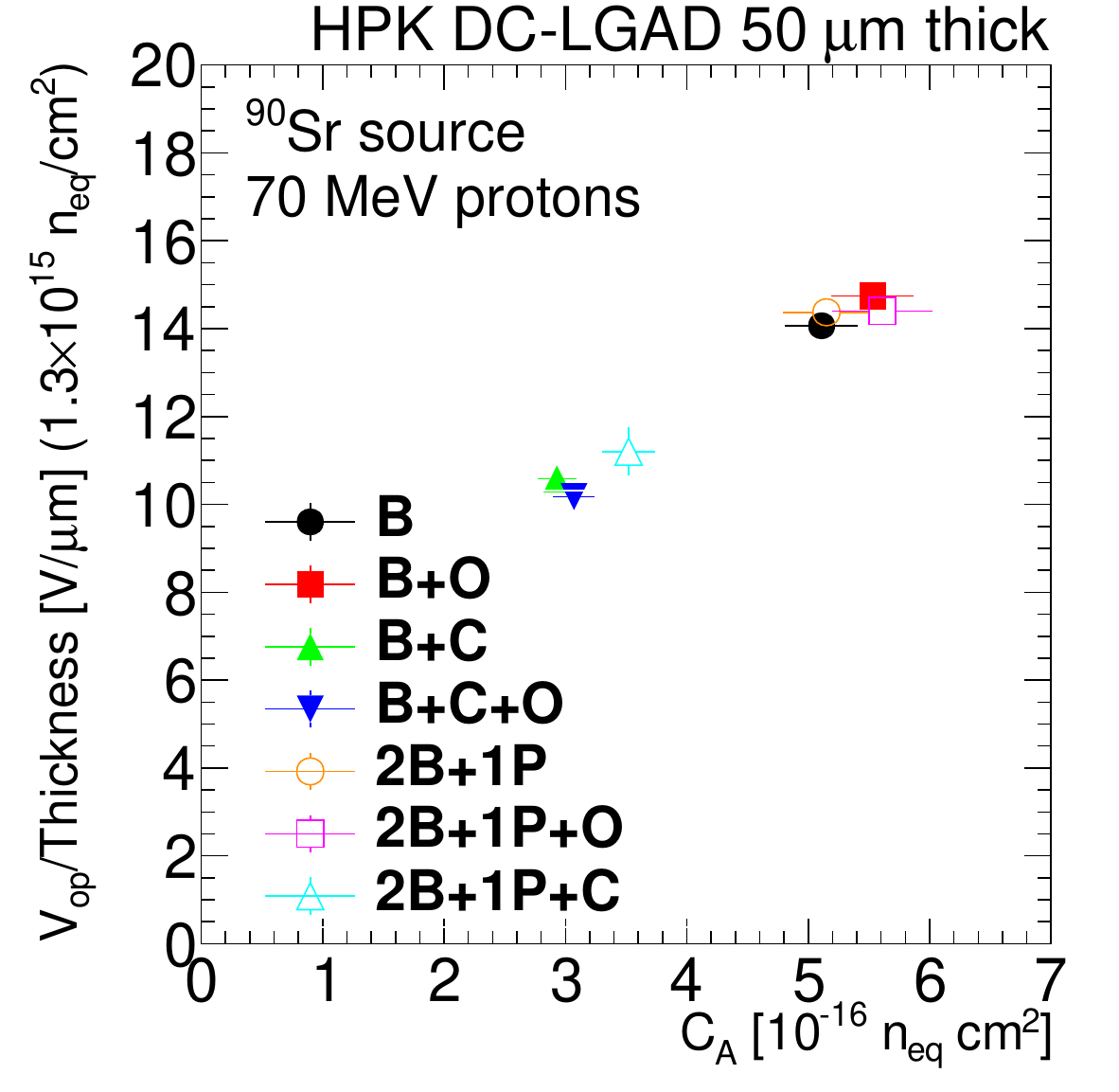}
\caption{Correlation between the acceptor-removal coefficient $C_{\mathrm{A}}$ and the operation voltage $V_{\mathrm{op}}$ measured at a fluence of $1.3\times10^{15}\,\neqcm$ for the different gain-layer designs. Samples with smaller $C_{\mathrm{A}}$ systematically require lower post-irradiation operation voltage.}
\label{fig:CA_vs_Vop}
\end{figure}

To compare the two radiation-tolerance observables directly, Fig.~\ref{fig:CA_vs_Vop} shows the relation between the acceptor-removal coefficient $C_{\mathrm{A}}$ extracted from the IV analysis and the operation voltage measured after irradiation at a fluence of $1.3\times10^{15}\,\neqcm$. A clear correlation is observed: samples with smaller $C_{\mathrm{A}}$ systematically require lower operation voltage after irradiation.

The reference sample B, the high-oxygen boron-only sample B+O and the compensation-only samples are clustered in the region of larger $C_{\mathrm{A}}$ and higher $V_{\mathrm{op}}$, while the carbon-implanted samples are shifted toward smaller $C_{\mathrm{A}}$ and lower $V_{\mathrm{op}}$. The combined sample 2B+1P+C is located in the same region as the other carbon-implanted samples within the present uncertainties. This correlation demonstrates that the reduction of $C_{\mathrm{A}}$ observed in the IV analysis is directly reflected in the operation voltage required to recover the timing performance after irradiation.

\subsection{Particle and energy dependence of acceptor removal}

\begin{figure}[tbp]
\centering
\includegraphics[width=0.85\linewidth]{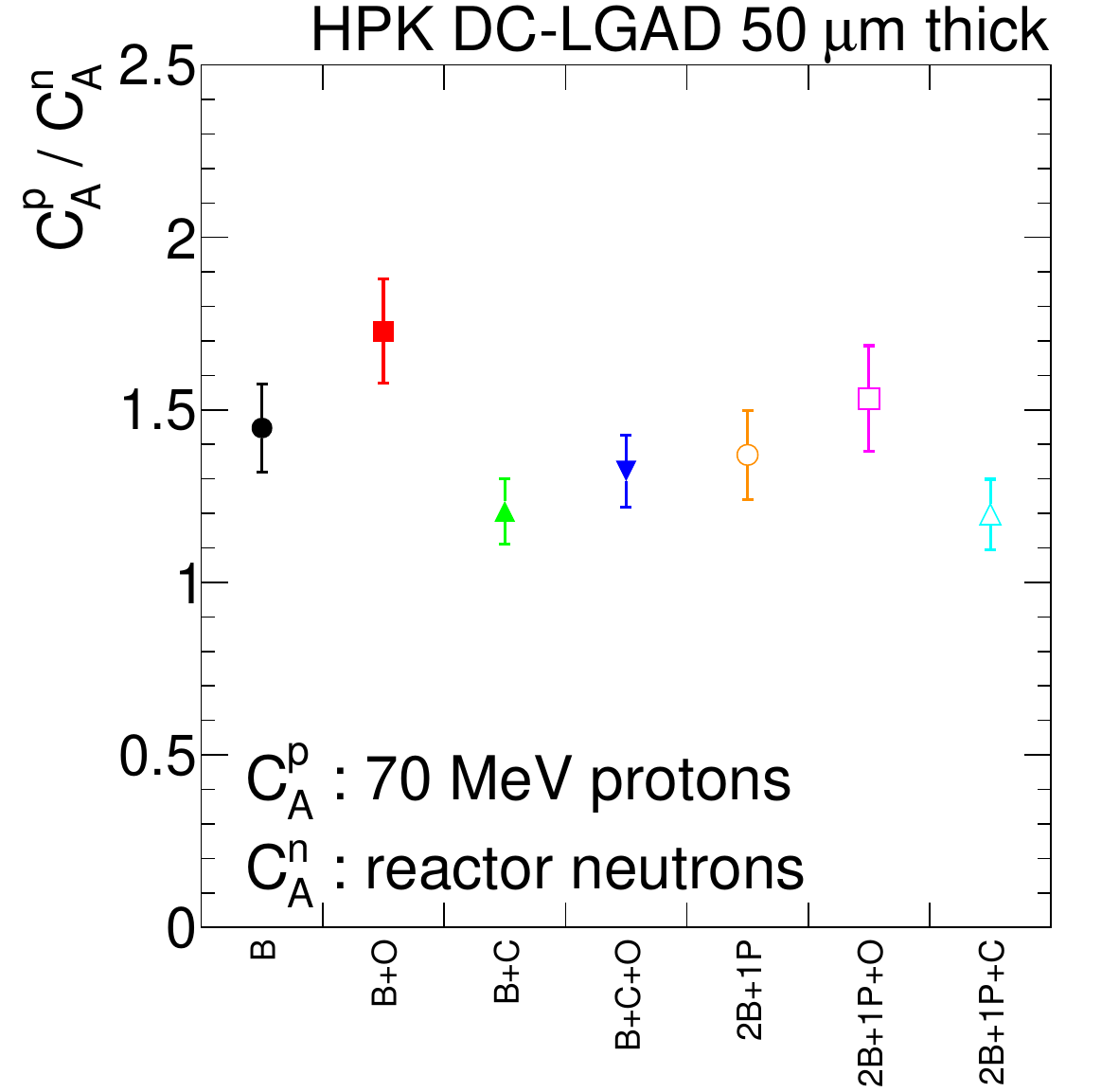}
\caption{Ratio of the acceptor-removal coefficients extracted from $70\,\mathrm{MeV}$ proton irradiation and reactor-neutron irradiation for the sample types studied in both campaigns.}
\label{fig:CpCn_ratio}
\end{figure}

A subset of the prototype sensors was also irradiated with reactor neutrons in order to compare the gain-layer degradation under different irradiation particles. The comparison is summarized here in terms of the extracted acceptor-removal coefficients. Table~\ref{tab:cA_pn} lists the values obtained for $70\,\mathrm{MeV}$ proton irradiation, denoted by $C_{\mathrm{A}}^{\mathrm{p}}$, and for reactor-neutron irradiation, denoted by $C_{\mathrm{A}}^{\mathrm{n}}$, together with their ratio. The ratio $C_{\mathrm{A}}^{\mathrm{p}}/C_{\mathrm{A}}^{\mathrm{n}}$ is shown in Fig.~\ref{fig:CpCn_ratio}.

\begin{table}[htbp]
\centering
\caption{Ratio $C_{\mathrm{A}}^{\mathrm{p}}/C_{\mathrm{A}}^{\mathrm{n}}$ of the acceptor-removal coefficients extracted from $70\,\mathrm{MeV}$ proton irradiation and reactor-neutron irradiation for the sample types studied in both campaigns.}
\label{tab:cA_pn}
\begin{tabular}{lccc}
\hline
\multirow{2}{*}{Sample}
& $C_{\mathrm{A}}^{\mathrm{p}}$
& $C_{\mathrm{A}}^{\mathrm{n}}$
& \multirow{2}{*}{$C_{\mathrm{A}}^{\mathrm{p}}/C_{\mathrm{A}}^{\mathrm{n}}$} \\
& [$10^{-16}\,\neqcm$]
& [$10^{-16}\,\neqcm$]
& \\
\hline
B         & $5.11 \pm 0.30$ & $3.53 \pm 0.23$ & $1.45 \pm 0.13$ \\
B+O       & $5.53 \pm 0.34$ & $3.20 \pm 0.20$ & $1.73 \pm 0.15$ \\
B+C       & $2.93 \pm 0.16$ & $2.43 \pm 0.14$ & $1.21 \pm 0.10$ \\
B+C+O     & $3.07 \pm 0.17$ & $2.32 \pm 0.13$ & $1.32 \pm 0.10$ \\
2B+1P     & $5.15 \pm 0.36$ & $3.76 \pm 0.24$ & $1.37 \pm 0.13$ \\
2B+1P+O   & $5.61 \pm 0.41$ & $3.66 \pm 0.25$ & $1.53 \pm 0.15$ \\
2B+1P+C   & $3.52 \pm 0.22$ & $2.94 \pm 0.17$ & $1.20 \pm 0.10$ \\
\hline
\end{tabular}
\end{table}

For the sample subset studied in both irradiations, the acceptor-removal coefficients extracted from reactor-neutron irradiation, $C_{\mathrm{A}}^{\mathrm{n}}$, are systematically smaller than those obtained from $70\,\mathrm{MeV}$ proton irradiation, $C_{\mathrm{A}}^{\mathrm{p}}$. The ratio $C_{\mathrm{A}}^{\mathrm{p}}/C_{\mathrm{A}}^{\mathrm{n}}$ is typically of the order of 1.2--1.7, depending on the sample type. A slight tendency toward smaller ratios may be present for the carbon-implanted samples, although no statistically significant sample-dependent deviation from the overall trend is resolved within the present uncertainties.

\begin{figure}[tbp]
\centering
\includegraphics[width=0.75\linewidth]{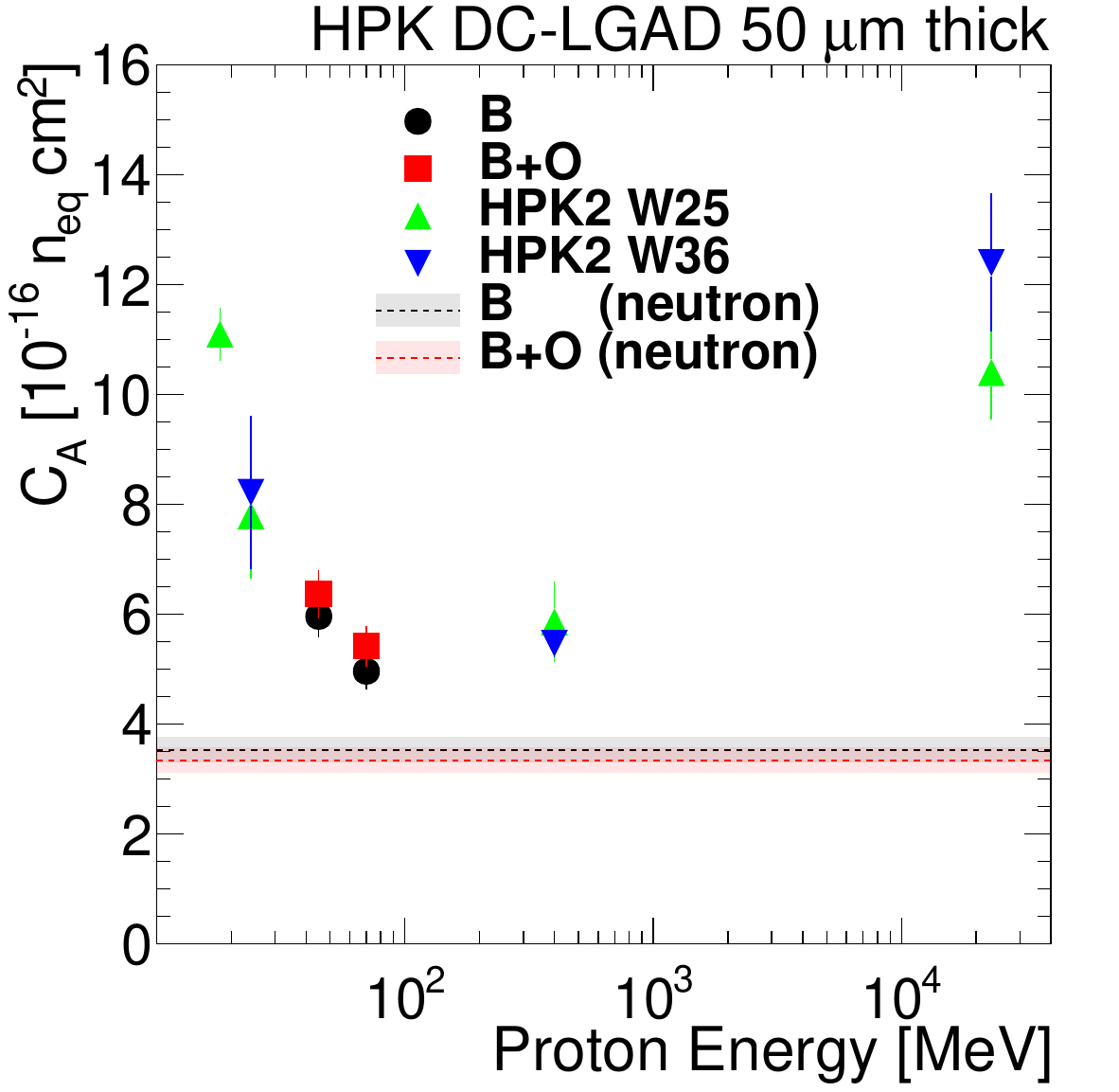}
\caption{Energy dependence of the acceptor-removal coefficient $C_{\mathrm{A}}$. The results obtained in this work for reactor neutrons, $45\,\mathrm{MeV}$ protons, and $70\,\mathrm{MeV}$ protons are compared with previously reported proton-irradiation data from the literature~\cite{Kraus:2954269}.}
\label{fig:CA_vs_proton_energy}
\end{figure}

Figure~\ref{fig:CA_vs_proton_energy} shows the energy dependence of the acceptor-removal coefficient $C_{\mathrm{A}}$ for proton irradiation. The proton results obtained in this work at $45\,\mathrm{MeV}$ and $70\,\mathrm{MeV}$ may therefore be viewed as $C_{\mathrm{A}}^{\mathrm{p}}(45\,\mathrm{MeV})$ and $C_{\mathrm{A}}^{\mathrm{p}}(70\,\mathrm{MeV})$, respectively, while the reactor-neutron results are shown separately for comparison.

\section{Discussion}

The results presented above show a clear overall trend. Among the approaches investigated in this work, carbon implantation is the only one that leads to a substantial improvement in radiation tolerance, whereas neither oxygen-related modification nor gain-layer compensation produces a clear improvement within the present measurement precision. In the following, possible physical interpretations of these observations are discussed.

\subsection{Role of carbon implantation in radiation-tolerance improvement}

The improvement observed for the carbon-implanted samples is qualitatively consistent with the commonly discussed competition picture for interstitial-driven defect reactions in irradiated LGAD gain layers. In this picture, displacement-induced silicon interstitials can react with either substitutional boron or substitutional carbon, i.e.
\[
\mathrm{B_s + Si_i \rightarrow B_i + Si_s},
\]
and
\[
\mathrm{C_s + Si_i \rightarrow C_i + Si_s}.
\]
If a significant fraction of the interstitial population is captured by carbon, the number of interstitials available to convert substitutional boron into electrically inactive configurations is reduced. In this way, carbon implantation can suppress the effective acceptor-removal rate.

The present results are consistent with this picture. Carbon implantation is the only modification that produces a clear reduction of both the acceptor-removal coefficient and the post-irradiation operation voltage. This consistency between the IV-based and timing-based observables supports the interpretation that carbon acts primarily by mitigating gain-layer degradation rather than by changing the initial effective gain-layer doping.

\subsection{Limited impact of oxygen reduction on gain-layer degradation}

One of the motivations of the present study was to test whether reducing the oxygen concentration in the gain layer could suppress acceptor removal by limiting the formation of boron--oxygen related defect complexes. However, no significant improvement is observed for either the low-oxygen samples or the PAB samples. Since the PAB approach was specifically intended to reduce the effective impact of oxygen-related reactions, its failure to improve the radiation tolerance provides additional evidence that oxygen contamination is not the dominant parameter controlling the macroscopic gain-layer degradation in the present HPK structures.

A plausible interpretation is that, in the highly doped LGAD gain layer, boron--oxygen related defects such as $\mathrm{B_iO_i}$ are not the dominant reaction channel for mobile interstitial boron. Ref.~\cite{kimerling1989} reported reaction branching ratios in p-type silicon indicating that the $\mathrm{B_iB_s}$ channel can be significantly stronger than the $\mathrm{B_iO_i}$ channel, with a ratio $[\mathrm{B_iB_s}]/[\mathrm{B_iO_i}] = 12$ under the conditions considered in their analysis. Their Table II also gives $[\mathrm{B_iO_i}]/[\mathrm{C_iO_i}] \approx [\mathrm{B_i}]/[\mathrm{C_i}] = 7$, which indicates that carbon-related channels can in principle compete with boron-related reactions, but also suggests that a substantial carbon concentration is required for this competition to become effective. In such a picture, simply reducing oxygen does not strongly modify the dominant acceptor-removal pathway, whereas sufficiently enhanced carbon-related reactions can. This is consistent with the absence of a clear improvement in the low-oxygen and PAB samples.

\subsection{Interpretation of the compensation results}
\subsubsection{Experimental inconsistency with the independent-removal picture}
The compensation approach was introduced with the expectation that donor removal could partially offset the reduction of the effective acceptor excess over donors. In the simplest independent-removal picture, if boron acceptors and phosphorus donors evolve independently under irradiation, the compensated gain layer should retain a larger effective doping than a purely boron-doped layer once carbon has reduced the acceptor-removal rate. In such a scenario, the combined sample 2B+1P+C would be expected to show better radiation tolerance than the carbon-only sample B+C.

Experimentally, however, this behavior is not observed. The compensation-only samples show radiation tolerance very similar to that of the boron-only references, and the combined sample 2B+1P+C does not show a clear improvement over B+C in either $C_{\mathrm{A}}$ or $V_{\mathrm{op}}$. These results suggest that a description based on independent acceptor and donor removal is incomplete for the present structures.

The simplest interpretation of the data is therefore that the evolution of the effective acceptor excess over donors in the compensated gain layer cannot be understood as a trivial superposition of an acceptor-removal term and a donor-removal term. Rather, the donor and acceptor channels may influence each other through the underlying displacement-induced defect kinetics. The present data do not identify the microscopic mechanism uniquely, but they do indicate that the compensation concept is less straightforward than expected from an independent-removal picture.

\subsubsection{Possible interplay between acceptor and donor removal}

A possible interpretation of the compensation results is that acceptor removal and donor removal are not independent because both processes are coupled through the common population of displacement-induced defects. In the standard picture, acceptor removal is driven mainly by interstitial-mediated reactions involving substitutional boron, while donor removal can be described by vacancy-related reactions involving substitutional phosphorus, for example,
\[
\mathrm{V + P_s \rightarrow VP},
\]
where $\mathrm{V}$ denotes a vacancy, $\mathrm{P_s}$ a substitutional phosphorus atom, and $\mathrm{VP}$ a vacancy--phosphorus complex. If the capture of vacancies by phosphorus changes the balance between vacancies and interstitials, it may also affect vacancy--interstitial recombination,

\[
\mathrm{V + Si_i \rightarrow Si_s},
\]
where $\mathrm{Si_i}$ and $\mathrm{Si_s}$ denote an interstitial silicon atom and a silicon atom on a regular lattice site, respectively. This reaction corresponds to the annihilation of a vacancy and an interstitial, and therefore modifies the effective interstitial population available to drive boron deactivation.

Within such a picture, phosphorus implantation could have two competing effects. On the one hand, donor removal may reduce the donor contribution to the effective space charge, which is the original motivation of compensation. On the other hand, vacancy capture into $\mathrm{VP}$-type defects may suppress vacancy--interstitial recombination and thereby increase the survival probability of interstitials. Since interstitials are the primary agents of boron deactivation, this could partially enhance acceptor removal and offset the benefit expected from donor removal.

At present, this interpretation remains a hypothesis. The present measurements establish the macroscopic inconsistency of the independent-removal picture, but they do not yet identify the specific microscopic defect reactions responsible for it. Additional studies, such as defect spectroscopy, annealing-dependent measurements, and simulations of defect reaction networks, would be required to test whether vacancy trapping by phosphorus or related mechanisms are indeed responsible for the observed behavior.

\section{Implications for LGAD gain-layer design}

The present results provide several practical guidelines for radiation-hard LGAD gain-layer design. First, carbon implantation should be regarded as the most promising design direction among the approaches studied here. It is the only modification that produces a clear and reproducible improvement in both the acceptor-removal coefficient and the post-irradiation operation voltage. At the same time, the non-irradiated measurements show that the present carbon-implanted structures exhibit larger leakage current and earlier breakdown. This indicates that future optimization should not aim simply at increasing the carbon concentration, but rather at reducing it as much as possible while preserving the improvement in acceptor removal. Such an optimization would be desirable in order to suppress the increase in leakage current and to avoid producing an excessive concentration of carbon-oriented defect complexes that behave as donors.

Second, oxygen reduction alone does not appear to be a particularly effective design handle for improving radiation tolerance in the present HPK structures. Although oxygen-related defect reactions are likely part of the microscopic defect network, the present results indicate that oxygen concentration is not the dominant parameter controlling the macroscopic gain-layer degradation in this doping regime. In practical terms, this suggests that lowering the oxygen concentration by itself is unlikely to yield a large gain in radiation tolerance.

Third, compensated gain-layer designs should be treated with caution. The present data show that compensation alone does not improve radiation tolerance, and that combining compensation with carbon does not provide a clear advantage over the carbon-only case. This implies that compensated structures cannot be optimized reliably on the basis of a simple independent-removal picture. Before compensation can be used as a robust design strategy, the underlying defect model should be tested experimentally through dedicated defect analyses, such as deep-level transient spectroscopy (DLTS), in order to clarify how donor-related and acceptor-related defect channels evolve under irradiation.

Finally, the observed difference between proton and neutron irradiations shows that the acceptor-removal coefficient depends on irradiation particle and energy. This has an important consequence for detector design studies: radiation tolerance should not be characterized by a single universal $C_{\mathrm{A}}$ value. More realistic performance projections require attention to the actual particle spectrum expected in operation, especially the contribution from low-energy hadrons.

\section{Conclusion}

In this work, we investigated the radiation tolerance of several HPK LGAD prototype sensors with modified gain-layer designs, including oxygen-modified, carbon-implanted, and boron--phosphorus compensated structures. Radiation tolerance was evaluated using two complementary observables: the acceptor-removal coefficient $C_{\mathrm{A}}$ extracted from IV measurements and the operation voltage $V_{\mathrm{op}}$ obtained from beta-ray timing measurements.

The oxygen-modified samples showed no significant improvement in radiation tolerance within the present uncertainties. In contrast, the carbon-implanted samples exhibited a clear reduction of $C_{\mathrm{A}}$ and a correspondingly lower operation voltage after irradiation. Compensation alone did not improve the radiation tolerance, and the combined compensated and carbon-implanted sample did not show a clear advantage over the carbon-only structures.

A comparison between proton and neutron irradiation further showed that the acceptor-removal coefficient is systematically smaller for reactor neutrons than for $70\,\mathrm{MeV}$ protons, and the present $45\,\mathrm{MeV}$ and $70\,\mathrm{MeV}$ proton data were found to be consistent with the general proton-energy dependence reported in the literature.

Taken together, these results identify carbon implantation as the only clearly effective method among those studied here for improving the radiation tolerance of the LGAD gain layer. Oxygen reduction and gain-layer compensation do not provide a significant benefit in the present HPK structures. The compensated samples also indicate that the evolution of the gain layer cannot be understood within a simple independent-removal picture.

\section*{Acknowledgments}
This research was partially supported by Grant-in-Aid for scientific research on advanced basic research (Grant No. 19H05193, 19H04393, 21H0073, 21H01099 and 25H00651) from the Ministry of Education, Culture, Sports, Science and Technology, of Japan as well as the Proposals for the U.S.-Japan Science and Technology Cooperation Program in High Energy Physics from JFY2019 to JFY2027 granted by High Energy Accelerator Research Organization (KEK) and Fermi National Accelerator Laboratory (FNAL). In conducting the present research program, the following facilities have been very important:  electron test beam provided at the AR test beamline (ARTBL) in KEK (Tsukuba) and Research Center for Accelerator and Radioisotope Science (RARiS) at Tohoku University.

\bibliographystyle{elsarticle-num-names}
\bibliography{ref}

\end{document}